\documentclass[11pt,journal,onecolumn,draftclsnofoot]{IEEEtran}

\usepackage[tbtags]{amsmath} % defines many math commands and
\usepackage{amssymb}  % get, among others, blackboard bold fonts
\usepackage{verbatim} % get comment environment, + new verbatim

\usepackage{enumerate}

\usepackage{amsfonts}
\usepackage{color}

\usepackage{amsthm}

\usepackage{cite}

\usepackage{epsfig, graphicx, psfrag}

\usepackage{soul}

\usepackage[cal=boondox]{mathalfa}
\providecommand{\Ber}{\mathcal{Ber}}
\providecommand{\Bin}{\mathcal{Bin}}

\newcommand{\khina}{{A.~Khina}}

\newcommand{\PI}{}
\newcommand{\Student}{}
\newcommand{\CoRes}{}

\newtheorem{thm}{Theorem}
\newtheorem{lemma}{Lemma}

\newtheorem{assert}{Assertion}
\newtheorem{corol}{Corollary}

\theoremstyle{definition}
\newtheorem{defn}{Definition}
\newtheorem*{defn*}{Definition}
\newtheorem{scheme}{Scheme}
\newtheorem*{scheme*}{Scheme}

\theoremstyle{remark}
\newtheorem{remark}{Remark}

\providecommand{\thmref}[1]{Th.~\ref{#1}}
\providecommand{\assertref}[1]{Assert.~\ref{#1}}

\providecommand{\secref}[1]{Sec.~\ref{#1}}
\providecommand{\lemref}[1]{Lem.~\ref{#1}}

\providecommand{\remref}[1]{Rem.~\ref{#1}}
\providecommand{\figref}[1]{Fig.~\ref{#1}}
\providecommand{\colref}[1]{Cor.~\ref{#1}}
\providecommand{\corolref}[1]{Cor.~\ref{#1}}

\providecommand{\schemeref}[1]{Sch.~\ref{#1}}

\newcommand{\ie}{i.e.}
\newcommand{\eg}{e.g.}

\newcommand{\etal}{et al.}

\newcommand{\nats}{\mathbb{N}}

\newcommand{\bm}[1]{\mbox{\boldmath{$#1$}}}

\newcommand{\e}{\text{e}}

\newcommand{\mG}{\mathcal{D}}
\newcommand{\mF}{\mathcal{E}}

\newcommand{\Comment}[1]{}
\newcommand{\old}[1]{}
\newcommand{\rem}[1]{}

\newcommand{\eps}{{\epsilon}}

\newcommand{\hs}{\hat{s}}

\providecommand{\tbs}{\tilde{\bs}}
\providecommand{\htbs}{\hat{\tbs}}
\providecommand{\ts}{\tilde{s}}
\providecommand{\hts}{\hat{\ts}}

\providecommand{\ts}{\tilde s}
\newcommand{\tS}{\tilde S}

\newcommand{\bs}{{\bm s}}
\newcommand{\tinyS}{{\footnotesize \bs}}

\newcommand{\by}{{\bm y}}

\providecommand{\hbs}{\hat{\bs}}

\newcommand{\bz}{{\bm z}}
\newcommand{\ba}{{\bm a}}

\providecommand{\be}{{\bm e}}
\newcommand{\bw}{{\bm w}}

\newcommand{\cN}{{\mathcal N}}

\newcommand{\Norm}[1]{\left\| #1 \right\|}

\providecommand{\e}{{\rm e}}
\providecommand{\comment}[1]{}

\providecommand{\norm}[1]{\Norm{#1}}

\newcommand{\beqn}[1]{\begin{eqnarray}\label{#1}}
\newcommand{\eeqn}{\end{eqnarray}}
\newcommand{\beq}[1]{\begin{equation}\label{#1}}
\newcommand{\eeq}{\end{equation}}

\providecommand{\var}[1]{{\rm Var\left( #1 \right)}}

\providecommand{\half}{\frac{1}{2}}
\providecommand{\half}{\frac{1}{2}}

\makeatletter
\newcommand{\vast}{\bBigg@{4}}
\newcommand{\Vast}{\bBigg@{5}}
\makeatother

\providecommand{\CMI}[3]{{I \left( #1 ; #2 \middle| #3 \right)}}

\providecommand{\h}[1]{{h \left( #1 \right)}}
\providecommand{\CH}[2]{{H \left( #1 \middle| #2 \right)}}
\providecommand{\Ch}[2]{{h \left( #1 \middle| #2 \right)}}

\providecommand{\oD}{\bar{D}}
\providecommand{\oR}{\bar{R}}

\providecommand{\E}[1]{\mathbb{E} \left[ #1 \right]}
\providecommand{\CE}[2]{\mathbb{E} \left[ #1 \middle| #2 \right]}
\providecommand{\Esub}[2]{\mathbb{E}_{#1} \left[ #2 \right]}
\providecommand{\CEsub}[3]{\mathbb{E}_{#1} \left[ #2 \middle| #3 \right]}

\providecommand{\packet}{\check{f}}
\providecommand{\PACKET}{f}
\providecommand{\EP}[1]{\cN\left( #1 \right)}
\providecommand{\CEP}[2]{\cN\left( #1 \middle| #2 \right)}
\providecommand{\IF}{}
\providecommand{\AND}{\textrm{ and }}

\providecommand{\RDF}[2]{\half \log \left( \frac{#1}{#2} \right)}

\providecommand{\Dweight}{D^\mathrm{Weighted}}
\providecommand{\betaut}{b}
\providecommand{\tind}{t}
\providecommand{\twopi}{2\pi}

\providecommand{\indind}{k}
\providecommand{\alphat}{\alpha_\tind}
\providecommand{\alphai}{\alpha_\indind}

\providecommand{\Wt}{W_\tind}

\providecommand{\Rt}{R_\tind}
\providecommand{\rt}{r_\tind}
\providecommand{\Ri}{R_\indind}
\providecommand{\Rpr}{R_{\indind-1}}
\providecommand{\Rone}{R_1}
\providecommand{\rone}{r_1}

\providecommand{\Npackets}{K}
\providecommand{\Recdq}{R} % R^\ECDQ}
\providecommand{\Decdq}{D} % D^\ECDQ}

\providecommand{\LQGcost}{\mathrm{J}}

\providecommand{\oLQGcost}{\bar{\LQGcost}}

\providecommand{\CostXs}{\mathsf{Q}}
\providecommand{\CostLastX}{\CostXs_T}
\providecommand{\CostUs}{\mathsf{R}}

\usepackage{ifthen}
\usepackage{mathtools}
\mathtoolsset{showonlyrefs=true} 

\usepackage[caption=false]{subfig}
\usepackage{hyperref}

\usepackage{tikz}
\usetikzlibrary{shapes,arrows,decorations.markings,positioning}

\begin{document}
%%%%%%%%%%%%%%%%%%%%%%%%%%%%%%%%%%%%%%%%%%%%%%%%%%%%%%%%%%
%%%%%%%%%%%%%%%%%%%%%%%%%%%%%%%%%%%%%%%%%%%%%%%%%%%%%%%%%%
%%%%%%%%%%%%%%%%%%%%%%%%%%%%%%%%%%%%%%%%%%%%%%%%%%%%%%%%%%

% \title{Sequential Coding of Gauss--Markov Sources \\ With Packet Erasures and Feedback}
% \title{Sequential Coding of Gauss--Markov Sources \\ over Packet-Erasure Channels with Feedback}
\title{Tracking and Control of Gauss--Markov Processes over Packet-Drop Channels with Acknowledgments}

\author{Anatoly Khina, Victoria Kostina, Ashish Khisti, and Babak Hassibi
	\thanks{This work was done, in part, while A.~Khina and V.~Kostina were visiting the Simons Institute for the Theory of Computing.
    This work has received funding from the European Union's Horizon 2020 research and innovation programme under the Marie Sk\l odowska-Curie grant agreement No 708932. 
    The work of V.~Kostina was supported in part by the National Science Foundation under Grant CCF-1566567.
    Ashish Khisti was supported by the Canada Research Chairs Program. 
	The work of B.~Hassibi was supported in part by the National Science Foundation under grants CNS-0932428, CCF-1018927, CCF-1423663 and CCF-1409204, by a grant from Qualcomm Inc., by NASA's Jet Propulsion Laboratory through the President and Director's Fund, by King Abdulaziz University, and by King Abdullah University of Science and Technology.
    The material in this paper was presented in part at the 2017 IEEE Information Theory Workshop.}
    \thanks{A.~Khina, V.~Kostina, and B.~Hassibi are with the Department of Electrical Engineering, California Institute of Technology, Pasadena, CA~91125, USA. \mbox{E-mails}: \mbox{{\em \{khina, vkostina, hassibi\}@caltech.edu}}}
    \thanks{A.~Khisti is with the Department of Electrical and Computer Engineering, University of Toronto, Toronto, ON M5S 3G4, Canada. E-mail: \mbox{\em akhisti@comm.utoronto.ca}}
}

\maketitle
%% To avoid page numbering
\thispagestyle{plain}
\pagestyle{plain}
%%%%%%%%%%%%%%%%%%%%%%%%%%%%%%%%%%%%%%%%%%%%%%%%%%%%%%%%%%%%%%%%%%%%%%%%%%%%%%%%%%%%%%%%%%%%%%%%%%%%%%%%%%

\begin{abstract}
    We consider the problem of tracking the state of Gauss--Markov processes over rate-limited  erasure-prone links.
    We concentrate first on the scenario in which several independent processes are seen by a single observer.
    The observer maps the processes into finite-rate packets that are sent over the erasure-prone links to a state estimator, and are acknowledged upon packet arrivals.
    The aim of the state estimator is to track the processes with zero delay and with minimum mean square error (MMSE).
    We show that, in the limit of many processes, 
    greedy quantization with respect to the squared error distortion is optimal. 
    That is, there is no tension between optimizing the MMSE of the process in the current time instant and that of future times. 
    For the case of packet erasures with delayed acknowledgments, 
    we connect the problem to that of compression with side information that is known at the observer and may be known at the state estimator~--- where the most recent packets serve as side information that may have been erased, and demonstrate that the loss due to a delay by one time unit is rather small.
    For the scenario where only one process is tracked by the observer--state estimator system, 
    we further show that variable-length coding techniques are within a small gap of the many-process outer bound. 
    We demonstrate the usefulness of the proposed approach for the simple setting of 
    discrete-time scalar linear quadratic Gaussian control with a limited data-rate feedback that is susceptible to packet erasures.
\end{abstract}


\allowdisplaybreaks
%%%%%%%%%%%%%%%%%%%%%%%%%%%%%%%%%%%%%%%%%%%%%%%%%%%%%%%%%%%%%%%%%%%%%%%%%%%%%%%%%%%%%%%%%%%%%%%%%%%%%%%%%%

\section{Introduction}
\label{s:intro}

Tracking the state of a system from noisy and possibly partially observable measurements 
is of prime importance in many estimation scenarios, and serves as an important building block in many control setups.

The recent rapid growth in wireless connectivity and its ad hoc distributed nature, 
while offering a plethora of new and exciting possibilities, 
introduces new design challenges for control over such media. 
These challenges include, among others, the need to track processes with minimal error over digital links of limited data rate 
which could be prone to (packet) erasures, 
and joint processing and reconstruction of distributed processes.

\begin{figure}[t]
\centering
	\includegraphics[width = \columnwidth, trim = 6.8cm 9.65cm 9.7cm 9.8cm]{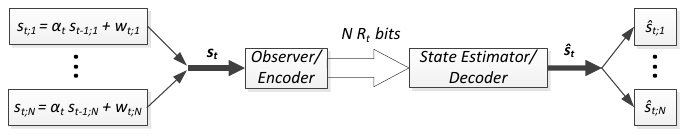}
    \caption{Multi-track of Gauss--Markov processes over a finite-rate channel.}
    \label{fig:multi_track}
\end{figure}

An important scenario, often encountered in practice, depicted in \figref{fig:multi_track},
is that of a \textit{multi-track} system that tracks several processes over a single shared  communication link.
In this scenario, at each time instant, 
several processes are observed by a single observer.
The observer, in turn, collects the measured states of these processes into a single vector state or \textit{frame},
and maps them into finite-rate packets, which are sent to the state-estimator 
over a channel which is prone to packet erasures.
The state estimator tracks the latest states of the different processes, by constructing minimum mean square error (MMSE) estimates thereof using the available packets received thus far.

Since these settings incorporate communication components, 
we appeal to relevant tools and results from information theory.
The information-theoretic framework for the multi-track setting with a large number of independent processes (large frames) and without packet erasures, 
was provided by Viswanathan and Berger~\cite{ViswanathanBerger:Streaming:IT} via the notion of \textit{sequential coding} 
for the case of two time steps, 
and for more steps in~\cite{MaIshwar:DPCM:IT,MaIshwar:DPCM:IT:ERRATUM,PredictiveAndCausalVideoCoding:Yang:IT1,PredictiveAndCausalVideoCoding:Yang:IT2}.
In these works the optimal tradeoff between given (per-process) rates and MMSEs (referred to  as \textit{distortions}) were determined when the number of processes is large, 
in the form of an optimization problem.

A similar framework in the context of control was also studied by Tatikonda~\cite{TatikondaPhD,BorkarMitterTatikonda:SequentialCoding,TatikondaSahaiMitter}, and Borkar \etal~\cite{BorkarMitterTatikonda:SequentialCoding} 
who noticed the intimate connection to the early works of Gorbunov and Pinsker \cite{GorbunovPinsker:CausalRDF,GorbunovPinsker:CausalRDF:Gauss}. Subsequent noteworthy efforts in the context of tracking include \cite{Tanaka:DirectedInfoGaussianRDF,Charalambous:NonanticipativeRDF-Filter:AC2014} and references therein.

For the special case of Gauss--Markov processes, 
an explicit expression for the achievable sum-rate for given distortions 
was derived in~\cite{MaIshwar:DPCM:IT,MaIshwar:DPCM:IT:ERRATUM} 
via the paradigms of predictive coding and differential pulse-code modulation (DPCM)~\cite{EliasPhD,Elias:PredictiveCoding,Oliver:PredictiveCoding,Harrison:PredictiveCoding,DPCM:Patent} (see also \cite[Ch.~6]{JayantNollBook} and the references therein),
and extended for the case of three time-steps of independent jointly Gaussian (not necessarily Markov) processes~--- in \cite{TorbatianYang:DPCM}.

In practice, packet-based protocols are prone to erasures and possible delays.
The multi-track scenario in the presence of packet erasures was treated under various erasure models. 
The case when only the first packet is prone to an erasure was considered in~\cite{SourceStreamingWithFirstPacketLoss}.
A more general approach that trades between the performance given all previously sent packets and the performance given only the last packet was proposed in \cite{GaussianRobustSequentialAndPredictiveCoding}. 
For random independent identically distributed (i.i.d.) packet erasures, a hybrid between pulse-code modulation (PCM) and DPCM, 
termed leaky DPCM was proposed in \cite{LeakyDPCM} and analyzed for the case of very low erasure probability in \cite{LeakyDPCM:HuangKochmanWornell}.
The scenario in which the erasures occur in bursts was considered in \cite{EtezadiKhistiTrott:IT,EtezadiKhistiChen:IT}. 

All of these works correspond to User Datagram Protocol (UDP) based networks~\cite{SchenatoSinopoliFranceschettiPoolaSSS}, 
in which no acknowledgment (ACK) of the arrival status of transmitted packets is available. That is, the observer does not know whether transmitted packets successfully arrived to the state estimator or were erased in the process.

In contrast, in Transmission Control Protocol (TCP) based networks, packet arrivals are acknowledged via a communication feedback link, in order to robustify the transmission of the overlying data~\cite{SchenatoSinopoliFranceschettiPoolaSSS}. Stabilizing control systems under this scenario has been studied in various works, \cite{LQRwithErasures,YukselMeyn:StabilzeMarkovoverErasures,Matveev:ControlInfoComputation:Springer
}, to name a few.

In this paper, we first consider the multi-track scenario of Gauss--Markov processes, which is defined formally in \secref{s:model}. 
We determine the optimal tradeoff between rates and distortions when the number of processes (frame length) is large, in \secref{s:no-drops}. 
Specifically, we show in \secref{s:no-drops} that greedy quantization that optimizes the distortion at each time is also optimal for minimizing the distortion of future time instants. 
This insight allows us to extend the result to the case where the compression rate $\rt$ available for the transmission of the packet at time $t$ is determined just prior to its transmission, in \secref{s:RandomRate}.

The packet-erasure channel with instantaneous ACKs can be viewed as a special case of the above noiseless channel with random rate allocation, with \mbox{$\rt = 0$} corresponding to a packet-erasure event~\cite{MineroFranceschettiDeyNair}.
The optimal tradeoff between rates and distortions for the multi-track scenario of Gauss--Markov processes in the presence of packet erasures and instantaneous ACKs thereby follows as a simple particularization of our more general result, as is shown in \secref{s:InstantFB} for both one-packet and multi-packet per state frame scenarios.

We further tackle, in \secref{s:Kaspi}, the more challenging delayed ACK setting, in which the observer does not know whether the most recently transmitted packets have arrived or not. By viewing these recent packets as side information (SI) that is available at the observer, and possibly at the state estimator, 
and leveraging the results of Kaspi~\cite{Kaspi94} along with their specialization for the Gaussian case by Perron \etal \cite{PerronDiggaviTelatar:GaussianKaspi:ISIT2006},\footnote{The scenario considered in \cite{Kaspi94,PerronDiggaviTelatar:GaussianKaspi:ISIT2006} can be also viewed as special case of the results of Heegard and Berger~\cite{HeegardBerger85}, where the SI is not available at the observer, by adjusting the distortion measure and ``augmenting'' the state~\cite{Causal-Noncausal_IEEEI2012}. Interestingly, knowing the SI at the observer allows one to improve the optimal performance of this scenario in the Gaussian case; see \remref{rem:SI@Tx_helps}.} we adapt our transmission scheme of \secref{s:no-drops} to the case of delayed ACKs.
We provide a detailed description of the proposed scheme for the case where ACKs are delayed by one time unit and demonstrate that the loss compared to the case of instantaneous ACKs is small.

In \secref{s:variable_rate}, we go on and consider the case of tracking a single process~---\textit{single-track}, 
and a variable-length coding (VLC) scenario~\cite{Huffman52}, \cite[Ch.~5]{CoverBook2Edition}, 
in which the packet size is not fixed and is instead constrained to be below a desired rate \textit{on average}. We consider a scheme that sequentially applies entropy-coded dithered quantization (ECDQ)~\cite{Ziv85,ZamirFeder:ECDQ,ZamirFeder:PrePostECDQ}, \cite[Ch.~5]{ZamirBook}, 
redolent of the scheme in~\cite{DerpichOstergaard:ECDQ4CausalRDF}, 
and show that it attains an MMSE--rate tradeoff that is close to the large-frame outer bound of \secref{s:no-drops}.

By supplementing the state tracking task with appropriate control actions 
in \secref{s:control}, 
we demonstrate the applicability of the derived results in Secs.~\ref{s:no-drops} and \ref{s:InstantFB}
to the scenario of  linear quadratic Gaussian (LQG) networked control, 
where a scalar linear plant driven by an i.i.d.\ Gaussian process is stabilized 
by a controller that is not co-located with the observer and is separated from it, instead, 
by a packeted communication (and more generally, a random-rate budget) channel.
We derive inner and outer bounds, on the optimal LQG cost
that extend those in~\cite{SilvaDerpichOstergaard:ECDQ4Control,KostinaHassibi:RDF4Control:AC} to packet-erasure channels.

We conclude the paper with \secref{s:discussion},
by discussing the cases of large delays, other types of VLC compression, 
and single-track with fixed-length coding (FLC) compression.

%---------------------------------------------------------------------------------------

\subsection{Notation}
\label{ss:notation}

Throughout the paper, $\norm{\cdot}$ denotes the Euclidean norm throughout this paper.
$\nats$ is the set of natural numbers. 
Random variables are denoted by lower-case letters with temporal subscripts 
($a_\tind, \hat{\tilde{a}}_\tind$), and random vectors (frames) of length $N \in \nats$ by boldface possibly accented lower-case letters ($\ba, \hat{\tilde{\ba}}_\tind$). % with temporal subscripts ($\ba_t, \hat{\tilde{\ba}}_\tind$). 
We denote temporal 
sequences by $\ba^t \triangleq \left( \ba_1, \ldots, \ba_t \right)$, 
where $\ba_t \triangleq \mathrm{Transpose} \left\{ \begin{pmatrix} a_{t;1} & a_{t;2} & \cdots & a_{t;N} \end{pmatrix} \right\}$,
% , with $t$ and $k$ used as the time indices.
and $[T] \triangleq \{1, \ldots, T\}$ is the interval from 1 to $T \in \nats$.
All other notations represent deterministic scalars.

%%%%%%%%%%%%%%%%%%%%%%%%%%%%%%%%%%%%%%%%%%%%%%%%%%%%%%%%%%%%%%%%%%%%%%%%%%%%%%%%%%%%%%%%%%%%%

% \vspace{-.5\baselineskip}
\section{Problem Statement}
\label{s:model}

We assume that t communication spans the time interval $[T]$ of horizon $T \in \nats$.

We next describe the state dynamics, and the operations carried by the observer and the state estimator, which communicate over a finite-rate channel, all of which are also depicted in \figref{fig:multi_track}.

\textit{State dynamics.}
Consider $N \in \nats$ independent Gauss--Markov processes $\{s_{\tind;1}\}$, $\{s_{\tind;2}\}$, \ldots, $\{s_{\tind;N}\}$ with identical statistics. This can be compactly represented in a vector form as (we assume $\bs_0 = 0$ for convenience):\footnote{The proposed treatment can be generalized to a matrix $\alpha$, but is much more involved and therefore remains outside the scope of this work.}
\begin{align}
\label{eq:s_t:recursion}
    \bs_\tind &= \alphat \bs_{\tind-1} + \bw_\tind, & \tind \in [T] \,,
\end{align}
where $\bs_t$ is the \textit{vector state} or \textit{frame} at time $t$,
$\{\alphat\}$ are known process coefficients,
the entries of $\bw_\tind$ are the $N$ independent driving noises, the entries of which are i.i.d.\ Gaussian with zero mean and variance~$\Wt$. We assume $\bs_0 = 0$ for convenience.

% \makebox[\linewidth]{Denote by $S_\tind$ the average power of $\bs_\tind$. Then,} 
Denote the average power of each state at time $t$ by
\mbox{$S_\tind \triangleq \E{s_{\tind;n}^2}$}, $n \in [N]$.
Then, \eqref{eq:s_t:recursion} implies 
the following recursive relation:
\begin{subequations}
\label{eq:P}
\noeqref{eq:P:recursion,eq:P:S0}
\begin{align}
    S_\tind &= \alphat^2 S_{\tind-1} + \Wt , & \tind \in [T] \,,
\label{eq:P:recursion}
 \\* S_0 &= 0 .
\label{eq:P:S0}
\end{align}
\end{subequations}

\textit{Observer.}
Sees the states $\{s_{\tind;1}, \ldots, s_{\tind;N}\}$ of all the $N$ process at time $\tind$, collects them into the frame $\bs_\tind$ and applies a causal function $\mF_\tind$ to the entire observed frame sequence $\bs^\tind$, to generate the packet $\PACKET_\tind \in \left[2^{N \Rt} \right]$:
\begin{align}
\label{eq:Encoder:Rule}
    \PACKET_\tind = \mF_\tind \left( \bs^\tind \right) ,
\end{align}
where $\Rt$ is the per-process \textit{rate} available for transmission over the channel at time $\tind$.

\textit{Channel.}
At time $\tind$, a packet $\PACKET_\tind \in \left[ 2^{N \Rt} \right]$ is sent over a noiseless channel of (per-process) finite rate $\Rt$.

\textit{State estimator.}
Applies a causal function~$\mG_\tind$ to the sequence of received packets $\PACKET^\tind$, to 
construct an estimate $\hbs_\tind$ of $\bs_\tind$, at time~$\tind$:
\begin{align}
\label{eq:hs_t:def}
    \hbs_\tind = \mG_\tind \left( \PACKET^\tind \right).
\end{align}

\textit{Distortion.}
The average mean-square error distortion (or MMSE) at time $\tind$ is defined as
\begin{align}
\label{eq:Dt:def}
    D_\tind \triangleq \frac{1}{N} \E{ \Norm{\bs_\tind - \hbs_\tind}^2 } .
\end{align}
% where $\norm{\cdot}$ denotes the Euclidean norm.

In the important special case of 
% an \emph{asymptotically stationary} process:
fixed parameters,
\begin{align}
\label{eq:model:SS}
	&
	\begin{aligned}
        \alphat &\equiv \alpha,
	 \\ \Wt &\equiv W, 
    \end{aligned}
    & \tind \in [T] \,,
\end{align}
the average process power converges to
\begin{align}
    S_\infty = \frac{W}{1 - \alpha^2} \,,
\end{align}
assuming $|\alpha| < 1$.
% We shall refer to such a process as \emph{asymptotically stationary}.
% 
% For asymptotically stationary source processes, 
In that case, by taking the rate-budget to be fixed too,
\begin{align}
\label{eq:model:SS:R}
	R_t &\equiv R, & \tind \in [T], 
\end{align}
we further define the 
% time-averaged (steady-state) distortion by the limit of the Ces\`aro means of $\{D_\tind\}$:
steady-state distortion (assuming the limit exists):
\begin{align}
\label{eq:D:SS}
%     D_\infty \triangleq \lim_{T \to \infty} \frac{1}{T} \sum_{\tind=1}^T D_\tind \,.
    D_\infty \triangleq \lim_{T \to \infty} D_\tind \,.
\end{align}

\begin{defn}[Distortion--rate region]
\label{def:RDR}
    The \emph{distortion--rate region} is the closure of all achievable distortion tuples $D^T \triangleq (D_1, \ldots, D_T)$ for a rate tuple $R^T \triangleq (R_1, \ldots, R_T)$, for any $N$, however large; its inverse is the \emph{rate--distortion region}.
\end{defn}

\begin{defn}[Average-stage rate and distortion]
\label{def:tot-rate}
	The average-stage rate and distortion are defined as 
    \begin{subequations}\noeqref{eq:totalRateBudgetConstraint}
	\begin{align}
    \label{eq:VLC:Constraint}
        \oR_T &\triangleq \frac{1}{T} \sum_{t=1}^T \Rt \,,
	\\ \label{eq:totalRateBudgetConstraint}
    	\oD_T &\triangleq \frac{1}{T} \sum_{t=1}^T D_t \,,
    \end{align}
    \end{subequations}
	respectively. 
    We further denote the steady-state average-stage rate and distortion by 
    \begin{subequations}
    \noeqref{eq:totalRateBudgetConstraint:SS:R}
    \begin{align}
    	\oR_\infty &= \limsup_{T \to \infty} \oR_T \,,
    \label{eq:totalRateBudgetConstraint:SS:R}
     \\ \oD_\infty &= \limsup_{T \to \infty} \oD_T \,.
    \label{eq:totalRateBudgetConstraint:SS}
    \end{align}
	\end{subequations}
%     Hence, a total rate constraint can be written 
%     as $\oR_T \leq R$.
\end{defn}

% \textbf{Goal:} Minimize~$\{ D_\tind \}$ for given $\{ \Rt \}$.

%%%%%%%%%%%%%%%%%%%%%%%%%%%%%%%%%%%%%%%%%%%%%%%%%%%%%%%%%%%%%%%%%%%%%%%%%%%%%%%%%%%%%%%%%%%%%%%%%%%%%%%%%%%

\section{Distortion--Rate Region of Gauss--Markov Process Multi-Tracking}
\label{s:no-drops}

The optimal achievable distortions for given
rates for the model of \secref{s:model}
are provided in the following theorem.

\begin{thm}[Distortion--rate region]
\label{thm:no-drop:RDR}
    The distortion--rate region of Gauss--Markov process multi-track for a rate tuple $R^T$ is given by all distortion tuples $D^T$ that satisfy $D_t \geq D^*_t$ with 
    \begin{subequations}
    \label{eq:RDR:Dt:general}
    \noeqref{eq:RDR:Dt:recursion:general,eq:RDR:Dt:D0:general}
    \begin{align}
        D^*_\tind &= \left( \alphat^2 D^*_{\tind-1} + \Wt \right) 2^{-2 \Rt}, & \tind \in [T] \,,
    \label{eq:RDR:Dt:recursion:general}
     \\ D^*_0 &= 0 .
    \label{eq:RDR:Dt:D0:general}
    \end{align}
    \end{subequations}
\end{thm}

\begin{remark}
	The impossibility (converse) of \thmref{thm:no-drop:RDR}
    has been established in \cite[Lem.~4.3]{TatikondaSahaiMitter}. 
    We provide an alternative simple proof in \secref{s:no-drops:OB} that allows us to treat random rates in the sequel.
\end{remark}

\begin{remark}
	The setting of \thmref{thm:no-drop:RDR} is referred to as ``causal encoder--causal decoder'' 
    by Ma and Ishwar~\cite{MaIshwar:DPCM:IT}. 
    We note that Ma and Ishwar~\cite{MaIshwar:DPCM:IT} provide an explicit result only for the sum-rate for the Gauss--Markov case~\cite{MaIshwar:DPCM:IT:ERRATUM}.{, where for the case of Gauss--Markov processes an explicit expression is provided only for the sum-rate.}
    Torbatian and Yang~\cite{TorbatianYang:DPCM} extend the sum-rate result to the case of 
    three-step general jointly Gaussian processes (which do not necessarily constitute a Markov chain).
    Our work, on the other hand, fully characterizes the rate--distortion region for the case of Gauss--Markov processes.
\end{remark}

\begin{remark}
    The results and proof (provided in the sequel) of \thmref{thm:no-drop:RDR} 
    imply that optimal greedy quantization at every step~--- which is achieved via Gaussian backward~\cite[Ch.~10.3]{CoverBook2Edition} or forward~\cite[pp.~338--339]{CoverBook2Edition} channels~--- becomes optimal when $N$ is large.
    Moreover, it achieves the optimum for all \mbox{$\tind \in [T]$} simultaneously, 
    meaning that there is no tension between minimizing the current distortion and future distortions.
\end{remark}

To prove \thmref{thm:no-drop:RDR} we first construct the optimal greedy scheme 
and determine its performance in \secref{s:no-drops:IB}.
We then show that it is in fact (globally) optimal when $N$ goes to infinity, by constructing an impossibility (outer) bound for this scenario, in \secref{s:no-drops:OB}.

%-----------------------------------------------------------------------------

\subsection{Achievability}
\label{s:no-drops:IB}

We construct an inner bound using the optimal greedy scheme, which amounts to the classical causal DPCM scheme.  % to be described next.
In this scheme all the quantizers are assumed to be MMSE quantizers, whose quantized 
values are well known to be uncorrelated with the resulting quantization errors. 

 \begin{scheme}[DPCM]%[No erasures]
\ 

\emph{Observer.} At time $\tind$:
    \begin{itemize}
    \item
    	Generates the prediction error
    	\begin{align}
    	\label{eq:no-drops:source2BeQuantized}
    	    \tbs_\tind \triangleq \bs_\tind - \alphat \hbs_{\tind-1} \,, 
    	\end{align}
    	where  
%         $\{\hbs_\tind | \tind = 1, \ldots, T\}$ 
		$\hbs_{\tind-1}$, defined in~\eqref{eq:hs_t:def},  is the previous frame reconstruction at the state estimator, and $\hbs_0 = 0$; a linear recursive relation for $\hbs_{\tind}$  is provided in the sequel in 
        \eqref{eq:no-drops:source_reconstruct}.\footnote{$\hbs_{\tind-1} = \CE{\bs_{\tind-1}}{\PACKET^{\tind-1}}$ and $\alphat \hbs_{\tind-1} = \CE{\bs_\tind}{\PACKET^{\tind-1}}$  
    	are the MMSE estimators of $\bs_{\tind-1}$ and $\bs_\tind$, respectively, 
    	given all outputs until time $t-1$.}
    \item
    	Generates $\htbs_\tind$, the quantized reconstruction of the prediction error $\tbs_\tind$, by quantizing $\tbs_\tind$
    	using the MMSE quantizer of rate $\Rt$ and frame length~$N$.
    \item
    	Sends $\PACKET_\tind = \htbs_\tind$ over the channel.
    \end{itemize}

\emph{State estimator.} At time $\tind$:
    \begin{itemize}
    \item
	Receives $\PACKET_\tind$.
    \item
        Recovers the reconstruction $\htbs_\tind$ of the prediction error $\tbs_\tind$.
    \item
	    Generates an estimate $\hbs_\tind$ of $\bs_\tind$:
	\begin{align}
	\label{eq:no-drops:source_reconstruct}
	    \hbs_\tind = \alphat \hbs_{\tind-1} + \htbs_\tind \,. 
	\end{align}
    \end{itemize}
\end{scheme}

\emph{Performance analysis.}
    First note that the error between $\bs_\tind$ and $\hbs_\tind$, denoted by $\be_\tind$, is equal to 
    \begin{subequations}
    \label{eq:general:UB:errors_relation}
    \noeqref{eq:general:UB:errors_relation:def,eq:general:UB:errors_relation:details,eq:general:UB:errors_relation:tildeS}
    \begin{align}
	    \be_\tind &\triangleq \bs_\tind - \hbs_\tind 
    \label{eq:general:UB:errors_relation:def}
     \\ &= \left( \tbs_\tind + \alphat \hbs_{\tind-1} \right) - \left( \alphat \hbs_{\tind-1} + \htbs_\tind \right)
    \label{eq:general:UB:errors_relation:details} 
     \\ &= \tbs_\tind - \htbs_\tind  \,,
        \label{eq:general:UB:errors_relation:tildeS}
    \end{align}
    \end{subequations}
    %and is of average power $D_\tind$, 
    where \eqref{eq:general:UB:errors_relation:details} follows from \eqref{eq:no-drops:source2BeQuantized} and \eqref{eq:no-drops:source_reconstruct}.
Thus, the distortion \eqref{eq:Dt:def} is also the distortion in reconstructing~$\tbs_\tind$.
    
    Using \eqref{eq:s_t:recursion}, \eqref{eq:no-drops:source2BeQuantized} and \eqref{eq:general:UB:errors_relation}, we express $\tbs_\tind$ as 
    \begin{align}
        \tbs_\tind &\triangleq \bs_\tind - \alphat \hbs_{\tind-1}
     \\ &= \alphat \left( \bs_{\tind-1} - \hbs_{\tind-1} \right) + \bw_\tind
     \\ & = \alphat \be_{\tind-1} + \bw_\tind \,. 
    \end{align}

    Since $\bw_\tind$ is independent of $\be_{\tind-1}$, the average power of the entries of $\tbs_\tind$ is equal to 
    \begin{align}
        \tS_\tind = \alphat^2 D_{\tind-1} + \Wt \,.
    \end{align}
    
    Using the property that the rate--distortion function under mean square error distortion of a process with a given average variance is upper bounded by that of an i.i.d.\ Gaussian process with the same variance (see, \eg, \cite[pp.~338--339]{CoverBook2Edition}), we obtain the following recursion: 
    \begin{align}
        D_\tind \leq \left( \alphat^2 D_{\tind-1} + \Wt \right) 2^{-2\Rt} ,
    \end{align}
    and hence \eqref{eq:RDR:Dt:general} is achievable within an arbitrarily small $\eps > 0$, for a sufficiently large $N$.
\hfill $\IEEEQED$

%-----------------------------------------------------------------------------

\subsection{Impossibility (Converse)}
\label{s:no-drops:OB}

    We now prove that, for any frame length $N \in \nats$, 
    \begin{subequations}
    \label{eq:ind2prove}
    \noeqref{eq:ind2prove:EP,eq:ind2prove:recur}
    \begin{align}
        D_\tind &\geq 2^{-2 \Rt} \Esub{\packet^{\tind-1}}{ \EP{\bs_\tind | \PACKET^{\tind-1} = \packet^{\tind-1}} } \ \ 
	\label{eq:ind2prove:EP}
     \\ &\geq D^*_\tind , \qquad\qquad \tind \in [T] \,,
	\label{eq:ind2prove:recur}
    \end{align}
    \end{subequations}
    by induction, 
    where the sequence $\{D_t^*\}$ is defined in~\eqref{eq:RDR:Dt:general}, 
    \begin{align}
    	\EP{\bs_\tind} &\triangleq \frac{1}{\twopi \e} 2^{\frac{2}{N} \h{\tinyS_\tind} }, 
	\\
    	\CEP{\bs_\tind}{\PACKET^\indind = \packet^\indind} & \triangleq \frac{1}{\twopi \e} 2^{\frac{2}{N} \Ch{\tinyS_\tind}{\PACKET^\indind = \packet^\indind} }
	\nonumber
    \end{align}
    denote the entropy-power (EP) and conditional EP of $\bs_\tind$ given $\PACKET^\indind = \packet^\indind$, 
	the expectation $\Esub{\packet^{\tind-1}}{\cdot}$ is with respect to $\packet^{\tind-1}$, 
	and the random vector $\packet^\tind$ is distributed the same as $\PACKET^\tind$.
%     and $\twopi = 2 \pi$ is the circle constant.

    \textbf{\emph{Basic step} ($t = 1$).}
    First note that, since $\bs_0 = 0$, and the vector $\bw_1$ consists of i.i.d.\ Gaussian entries of variance $W_1$, 
    \eqref{eq:ind2prove:recur} is satisfied with equality.
    To prove \eqref{eq:ind2prove:EP}, we use the fact that 
    the optimal achievable distortion $D_1$ for a Gaussian process ($\bs_1 = \bw_1$) with i.i.d.\ entries of power $W_1$ and rate $\Rone$ is dictated by its rate--distortion function~\cite[Ch.~10.3.2]{CoverBook2Edition}:
    \begin{align}
        D_1 \geq W_1 2^{-2 \Rone} .
    \end{align}
    
    \emph{\textbf{Inductive step.}}
    Let $\indind \geq 2$ and suppose \eqref{eq:ind2prove} is true for $\tind = \indind-1$.
    We shall now prove that it holds also for $\tind = \indind$.
    
    {{%\small
%     \vspace{-\baselineskip}
    \begin{subequations}
    \noeqref{eq:LB:ineq:D:def}
    \begin{align}
        D_\indind &=
     \frac{1}{N} \E{ \CE{ \Norm{ \bs_\indind - \hbs_\indind }^2 }{ \PACKET^{\indind-1} } } 
    \label{eq:LB:ineq:smoothing}
	 \\ &= \frac{1}{N} \Esub{\packet^{\indind-1}}{ \CE{ \Norm{ \bs_\indind - \hbs_\indind }^2 }{ \PACKET^{\indind-1} = \packet^{\indind-1} } } 
    \label{eq:LB:ineq:f2F}
 	 \\ &\geq \Esub{\packet^{\indind-1}}{ \CEP{\bs_\indind}{\PACKET^{\indind-1} = \packet^{\indind-1}} 2^{-2 \Ri} } 
    \label{eq:LB:ineq:SLB}
     \\ &= \Esub{\packet^{\indind-1}}{ 
     \EP{ \alphai \bs_{\indind-1} + \bw_\indind | \PACKET^{\indind-1} = \packet^{\indind-1}}
     } 2^{-2 \Ri} \quad
    \label{eq:LB:ineq:recursion}
     \\ &\geq \Big\{
        \Esub{\packet^{\indind-2}}{ \CEsub{\packet_{\indind-1}}{ \EP{\alphai \bs_{\indind-1} | \PACKET^{\indind-1} = \packet^{\indind-1}} }{ \packet^{\indind-2}} }
       + \EP{\bw_\indind} \Big\} 2^{-2\Ri} 
    \label{eq:LB:ineq:EPI}
     \\ &\geq \Big\{
        \alphai^2 \Esub{\packet^{\indind-2}}{ \CEP{ \bs_{\indind-1}}{\PACKET^{\indind-2} = \packet^{\indind-2},\, \PACKET_{\indind-1}} }
      + W_k \Big\} 2^{-2\Ri}
    \label{eq:LB:ineq:Jensen}
     \\ &\geq \Big\{
        \alphai^2  \Esub{\packet^{\indind-2}}{ \EP{\bs_{\indind-1} | \PACKET^{\indind-2} = \packet^{\indind-2}} } 2^{-2 \Rpr}
      + W_k \Big\} 2^{-2\Ri} 
    \label{eq:LB:ineq:Entropies}
     \\ &\geq 2^{-2\Ri} \left(
        \alphai^2 D^*_{\indind-1}
        + W_k \right) 
    \label{eq:LB:ineq:Hypothesis}
     \\  &= D^*_\indind ,
    \label{eq:LB:ineq:D*}
    \end{align}
    \end{subequations}
    }} % END \small
    
    \noindent where 
    \eqref{eq:LB:ineq:smoothing} follows from the law of total expectation,
    \eqref{eq:LB:ineq:f2F} holds since $\PACKET^{k-1}$ and $\packet^{k-1}$ have the same distribution,
    \eqref{eq:LB:ineq:SLB} follows by bounding from below the inner expectation (conditional distortion) by the rate--distortion function and  the Shannon lower bound~\cite[Ch.~10]{CoverBook2Edition} --- this also proves \eqref{eq:ind2prove:EP}, 
    \eqref{eq:LB:ineq:recursion} is due to \eqref{eq:s_t:recursion}, 
    \eqref{eq:LB:ineq:EPI} follows from the entropy-power inequality~\cite[Ch.~17]{CoverBook2Edition}, 
    \eqref{eq:LB:ineq:Jensen} holds since $\bw_\indind$ is Gaussian, 
    the scaling property of differential entropies
    and 
    Jensen's~inequality:
    \begin{subequations}
    \label{eq:LB:general:EP:Jensen}
    \begin{align}
        \CEsub{\packet_{\indind-1}}{ 2^{\frac{2}{N} \Ch{\tinyS_{\indind-1}}{\PACKET^{\indind-1} = \packet^{\indind-1}} } }{\packet^{\indind-2}}
    \nonumber
     &\geq 2^{\frac{2}{N} \Esub{\packet_{\indind-1}}{ \Ch{\tinyS_{\indind-1}}{\PACKET^{\indind-1} = \packet^{\indind-1}} }} 
     \\* &
     \equiv 2^{\frac{2}{N} \Ch{\tinyS_{\indind-1}}{\PACKET^{\indind-2} = \packet^{\indind-2},\, \PACKET_{\indind-1}} } \!,
    \label{eq:LB:general:Jensen:CondEntropy}
    \end{align}
    \end{subequations}
    \eqref{eq:LB:ineq:Entropies} follows from the following standard set of inequalities:
    \begin{align}
        N \Rpr & \CH{\PACKET_{\indind-1}}{\PACKET^{\indind-2} = \packet^{\indind-2}}
     \\ &\geq \CMI{\bs_{\indind-1}}{\PACKET_{\indind-1}}{\PACKET^{\indind-2} = \packet^{\indind-2}}
     \\ &= \Ch{\bs_{\indind-1}}{\PACKET^{\indind-2} = \packet^{\indind-2}} 
     	- \Ch{\bs_{\indind-1}}{\PACKET^{\indind-2} = \packet^{\indind-2}, \PACKET_{\indind-1}} ,
	\nonumber
    \end{align}
    \eqref{eq:LB:ineq:Hypothesis} is by the induction hypothesis, 
    and \eqref{eq:LB:ineq:D*} holds by the definition of $\{D^*_\tind\}$ as the sequence that satisfies \eqref{eq:RDR:Dt:general}~--- which also proves \eqref{eq:ind2prove:recur}.
    This concludes the proof of \eqref{eq:ind2prove:recur}.
    \hfill $\IEEEQED$

\begin{assert}[Outer bound for non-Gaussian noise]
\label{thm:LB:nonGaussian}
    Consider the setting of \secref{s:model} with independent non-Gaussian noise entries $\{w_{\tind;n} | \tind \in [T], n \in [N] \}$.
    Then, the average achievable distortion $D_\tind$ at time $\tind \in [T]$ is bounded from below by $D_\tind \geq D^*_\tind$, with $D^*_\tind$ given by the recursion 
    \begin{align}
    	D^*_\tind &= \left( \alpha^2 D^*_{\tind-1} + \EP{\bw_t} \right) 2^{-2\Rt} .
	 \\ D^*_0 &= 0 .
    \end{align}
\end{assert}

\begin{IEEEproof}
	The proof is identical to that of the lower bound for the Gaussian case with $W_t$ replaced by $\EP{\bw_t}$.\footnote{Recall that in the Gaussian setting $\EP{\bw_t} = \var{{\bw_t}}/N \equiv W_t$.}
\end{IEEEproof}

%-----------------------------------------------------------------------------

\subsection{Fixed-Parameter Gauss--Markov Processes}
\label{s:no-drops:SS}

For the case of fixed parameters~\eqref{eq:model:SS} and fixed rate, the steady-state average distortion is as follows.
\begin{corol}[Steady state performance with fixed-rate budget]
\label{corol:no-drops:SS}
    Assume a fixed-parameter~\eqref{eq:model:SS} fixed-rate budget~\eqref{eq:model:SS:R} setting. If $\alpha^2 2^{-2R} < 1$,\footnote{This is trivial for $|\alpha|<1$.} % (trivial for $|\alpha|<1$), 
    then the steady-state distortion is given by
    \begin{align}
    \label{eq:general:SS:D}
        D_\infty^* \triangleq \lim_{\tind \to \infty} D^*_\tind = \frac{W 2^{-2R}}{1 - \alpha^2 2^{-2R}} \,, 
    \end{align}
    and is otherwise unbounded.
\end{corol}

\begin{IEEEproof}
	The proof is immediate by noting that~\eqref{eq:general:SS:D} constitutes a linear time-invariant (LTI) system and therefore is globally exponentially stable if the (only) pole of its transfer function lies strictly inside the unit circle, \ie, $\alpha^2 2^{-2R} < 1$, and is unstable otherwise~\cite[Ch.~6]{LinearSystemsBook}.
    We provide a proof for completeness.
	Assume $\alpha^2 2^{-2R} < 1$. Then, \eqref{eq:general:SS:D} is a fixed point of \eqref{eq:RDR:Dt:recursion:general}.

    We now prove that 
    $D^*_\tind$ converges to $D^*_\infty$.
    Assume $D^*_{\tind-1} \neq D^*_\infty$ (otherwise we are already at the fixed point). Then, 
    \begin{align}
        D^*_\tind \!- D^*_\infty 
        &= \left[ \left( \alpha^2 D^*_{\tind-1} + W \right) 2^{-2R} \right] 
         - \left[ \left( \alpha^2 D^*_\infty + W \right) 2^{-2R} \right] 
	\nonumber
    \\* & = \alpha^2 2^{-2R} \left( D^*_{\tind-1} - D^*_\infty \right) , 
    \end{align}
    or equivalently
    \begin{align}
        \frac{D^*_\tind - D^*_\infty}{D^*_{\tind-1} - D^*_\infty} = \alpha^2 2^{-2R} < 1 .
    \end{align}
    Hence, if $0 \lessgtr D^*_{\tind-1} - D^*_\infty$, then 
    \begin{align}
        0 \lessgtr D^*_\tind - D^*_\infty \lessgtr D_{\tind-1} - D^*_\infty ,
    \end{align}
    meaning that $D^*_t$ converges (exponentially fast) to $D^*_\infty$.
    
    To prove the converse, assume $\alpha^2 2^{-2R} \geq 1$. Then, 
    \begin{align}
    	D_t^* &\geq D_{t-1}^* + \EP{\bw_t} 
	 \\ &\geq t \EP{\bw_t} , 
    \end{align}
    which goes to infinity for $t \to \infty$.
\end{IEEEproof}

\begin{remark}
    As is evident from the proof, the result of \corolref{corol:no-drops:SS}
    remains true for any initial value $D^*_0$.
\end{remark}

\begin{remark}
	The impossibility part of \colref{corol:no-drops:SS}
    can be traced back to the work of Gorbunov and Pinsker~\cite{GorbunovPinsker:CausalRDF:Gauss}.
\end{remark}

Interestingly, the optimal steady-state distortion achievable with a fixed-rate budget~\eqref{eq:model:SS:R} is in fact optimal even if we loosen this restriction to a total rate-budget constraint as was previously observed, \eg, 
in \cite{DerpichOstergaard:ECDQ4CausalRDF}.
This is a simple corollary of \thmref{thm:no-drop:RDR} and is formally proved next.
The same conclusion holds if the frame entries are correlated Gaussians, 
as was recently proved by Tanaka~\cite{Tanaka:UniformRateAllocation}. 

\begin{corol}[Steady state performance with total-rate budget]
\label{corol:no-drops:SS:total_budget}
    The average-stage steady-state distortion~\eqref{eq:totalRateBudgetConstraint:SS} $\oD_\infty$, 
    under a total rate-budget constraint~\eqref{eq:totalRateBudgetConstraint:SS:R} $\oR_\infty \leq R$, 
    is bounded from below by $\oD_\infty \geq D^*_\infty$.
    Consequently, the fixed (a.k.a.\ uniform) rate allocation $\Rt \equiv R$ is optimal in the limit of $T \to \infty$.    
\end{corol}

\begin{IEEEproof}
% \textit{Proof:}
	W.l.o.g., for a given tuple $R^T$, it suffices to consider distortion tuples $D^T$ that belong to the boundary of the rate--distortion region, namely, distortion tuples satisfying~\eqref{eq:RDR:Dt:recursion:general} with equality:
    \begin{align}
    \label{eq:RDR:Rt:recursion}
        R_t = \half \log \left( \alpha^2 D_{t-1} + W \right) - \half \log D_t \,.
    \end{align}

	For the equivalent problem of minimizing the total rate budget~\eqref{eq:VLC:Constraint} under an average-stage distortion constraint $\oD_T \leq D$, the total rate budget can be bounded from below as 
    \begin{subequations}
    \label{eq:uniformBudget}
   \noeqref{eq:uniformBudget:def,eq:uniformBudget:subst,eq:uniformBudget:rewrite,eq:uniformBudget:Jensen,eq:uniformBudget:constraint}
    \begin{align}
    \!\!\!
    	\oR_T &\equiv \frac{1}{T} \sum_{t = 1}^T R_t 
    \label{eq:uniformBudget:def}
	 \\ &= \frac{1}{T} \sum_{t=1}^T \left[ \half \log \left( \alpha^2 D_{t-1} + W \right) 
        - \half \log D_t \right]
    \label{eq:uniformBudget:subst}
     \\ &= \sum_{t=1}^T \frac{1}{2 T} \log \left( \alpha^2 + \frac{W}{D_t} \right) 
	 	- \frac{1}{2T} \log \left( 1 + \frac{\alpha^2 D_T}{W} \right) \quad\ \ 
    \label{eq:uniformBudget:rewrite}
     \\ &\geq \half \log \left( \alpha^2 + \frac{W}{\oD_T} \right)
        - \frac{1}{2T} \log \left( 1 + \frac{\alpha^2 T \oD_T}{W} \right)
    \label{eq:uniformBudget:Jensen}
     \\ &\geq 
     \half \log \left( \alpha^2 + \frac{W}{D} \right) 
       - \frac{1}{2T} \log \left( 1 + \frac{\alpha^2 T D}{W} \right) ,
    \label{eq:uniformBudget:constraint}
    \end{align}
    \end{subequations}
    where we use the definition of $\oR_T$~\eqref{eq:VLC:Constraint} in \eqref{eq:uniformBudget:def}, 
    \eqref{eq:uniformBudget:subst} holds by substituting~\eqref{eq:RDR:Rt:recursion}, 
    \eqref{eq:uniformBudget:Jensen} follows from Jensen's inequality and 
    \begin{align}
    	D_1 \leq \sum_{t=1}^T D_t \equiv T \oD_T ,
    \end{align}
    and \eqref{eq:uniformBudget:constraint} holds due to the constraint $\oD_T \leq D$.
    
    Evaluating \eqref{eq:uniformBudget} in the limit of 
    $T \to \infty$ 
    concludes the proof.
\end{IEEEproof}

%%%%%%%%%%%%%%%%%%%%%%%%%%%%%%%%%%%%%%%%%%%%%%%%%%%%%%%%%%%%%%%%%%%%%%%%%%%%%%%%%%%%%%%%%%%%%%%%%%%%%%%%%%%

\section{Random-Rate Budgets}
\label{s:RandomRate}

In practice, the available transmission rate may vary across time depending on the quality of service offered by the infrastructure, as well as, due to other applications sharing the same infrastructure.
We therefore generalize next the results of \secref{s:no-drops} to random rates $\{\rt\}$ 
that are independent of each other and of $\{\bw_t\}$. 
Due to the dynamic nature of the problem, the rate $\rt$ is revealed to the observer just before the transmission of time~$t$.

\begin{thm}[Distortion--rate region]
\label{thm:IF:RDR:Dt}
    The distortion--rate region of Gauss--Markov multi-track with
    independent rates $r^T$ is given by all distortion tuples $D^T$ that satisfy $D_t \geq D^*_t$ with 
	\begin{subequations}
	\label{eq:RDR:IF:Dt}
	\noeqref{eq:RDR:IF:Dt:recur,eq:RDR:IF:Dt:t=0}
    \begin{align}
	    D_\tind^* &= \left( \alphat^2 D_{\tind-1}^* + \Wt \right) \E{2^{-2 \rt}} , & \tind \in [T] 	
\,, \quad
	\label{eq:RDR:IF:Dt:recur}
	\\* D_0^* &= 0 .
	\label{eq:RDR:IF:Dt:t=0}
    \end{align}
    \end{subequations}
\end{thm}

\begin{IEEEproof}
\ 
    
    \textit{Achievability.}
    Since the achievability scheme in \thmref{thm:no-drop:RDR} does not use the knowledge of future transmission rates to encode and decode the packet at time $t$, we  have
    
    \begin{subequations}
    \label{eq:ConditionedDt}
    \noeqref{eq:ConditionedDt:def,eq:ConditionedDt:ineq,eq:ConditionedDt:causality}
    \begin{align}
    \\[-2.\baselineskip]
        d_\tind &\triangleq \frac{1}{N} \CE{ \Norm{\bs_\tind - \hbs_\tind}^2}{r^T}
    \label{eq:ConditionedDt:def}
     \\ &= \frac{1}{N} \CE{ \Norm{\bs_\tind - \hbs_\tind}^2}{r^\tind}
    \label{eq:ConditionedDt:causality}
    \\* &\leq (\alphat^2 d_{\tind-1} + \Wt) 2^{-2\rt} + \eps ,
    \label{eq:ConditionedDt:ineq}
    \end{align}
    \end{subequations}
    for any $\eps > 0$, however small, and large enough $N$.
    
    Taking an expectation of \eqref{eq:ConditionedDt:ineq} with respect to $r^t$ and using the independence of $r^{t-1}$ and $r_t$, we obtain \eqref{eq:RDR:IF:Dt}. 
    
    \textit{Impossibility.}
	Revealing the rates to the observer and the state estimator prior to the start of transmission can only improve the distortion. 
    Thus, the distortions $\{d_t\}$ conditioned on $\{\rt\}$~\eqref{eq:ConditionedDt:def} are bounded from below as in \thmref{thm:no-drop:RDR};
    by taking the expectation with respect to $\{\rt\}$, we attain the desired result.
\end{IEEEproof}

\begin{remark}
	By applying Jensen's inequality to \eqref{eq:RDR:IF:Dt:recur}:
    $\E{2^{-2 \rt}} \geq 2^{-2 \E{\rt}}$, 
	we see that using packets of a fixed rate equal to $\E{\rt}$ performs better than using random rates~$\rt$.
\end{remark}    

%-----------------------------------------------------------------------------------------

For the special case of fixed-parameters~\eqref{eq:model:SS} and i.i.d.\ rates $\{\rt\}$,
the steady-state distortion is given as follows.
\begin{corol}[Steady state]
\label{corol:RandomRate:SS}
	Assume a fixed-parameter~\eqref{eq:model:SS} setting with i.i.d.\ rates $\{r_t\}$.
    If $\alpha^2 B < 1$,\footnote{Again, this is trivial for $|\alpha|<1$.} 
    where $B \triangleq \E{2^{-2\rone}}$, 
    then the steady-state distortion is given by
    \begin{align}
    \label{eq:RandomRate:SS:D}
        D^*_\infty &\triangleq \lim_{\tind \to \infty} D^*_\tind = \frac{B W}{1 - \alpha^2 B} \,,
    \end{align}
    and is otherwise unbounded.
\end{corol}

\begin{IEEEproof}
	The proof is identical to that of~\colref{corol:no-drops:SS} with $2^{-R}$ replaced by $B$. 
\end{IEEEproof}

%%%%%%%%%%%%%%%%%%%%%%%%%%%%%%%%%%%%%%%%%%%%%%%%%%%%%%%%%%%%%%%%%%%%%%%%%%%%%%%%%%%%%%%%%%%%%%%%%%%%%%%%%%

\section{Packet Erasures with Instantaneous ACKs}
\label{s:InstantFB}

%--------------------------------------------------------------------------------------------------------

\subsection{One Packet Per Frame}
\label{ss:InstantFB:1packet}

An important scenario encompassed by the random-rate budget channel model of \secref{s:RandomRate} is that of packet erasures~\cite{MineroFranceschettiDeyNair}. 
Since a packet erasure at time $t$ can be viewed as $\rt = 0$, 
and assuming that the observer 
sends packets of fixed rate $R$ and  
is cognizant of any packet erasures instantaneously, 
the packet erasure channel can be cast as the random rate channel of \secref{s:RandomRate} with
\begin{subequations}
\label{eq:packet_erasures}
\noeqref{eq:packet_erasures:b_t,eq:packet_erasures:explicit}
\begin{align}
	\rt &= b_t R
\label{eq:packet_erasures:b_t}
 \\ &=
    \begin{cases}
    	R, & b_t = 1
     \\ 0, & b_t = 0
    \end{cases}
\label{eq:packet_erasures:explicit}
\end{align}
\end{subequations}
\begin{align}
\label{eq:packet_erasures}
	\rt &= R \, b_t
 	=
    \begin{cases}
    	R, & b_t = 1
     \\ 0, & b_t = 0
    \end{cases}
\end{align}
where $\{b_t\}$ are the packet-erasure events, such that $b_t = 1$ corresponds to a successful arrival of the packet $\PACKET_t$ at time $t$, and $b_t = 0$ means it was erased. We further denote by 
\begin{align}
\label{eq:received_packet}
	g_t \triangleq b_t \PACKET_t
\end{align}
the received output where $g_t = 0$ corresponds to an erasure, and otherwise $g_t = \PACKET_t$.
We assume that $\{b_t\}$ are i.i.d.\ according to a $\Ber(\beta)$ distribution for $\beta \in [0,1]$.

\begin{remark}
	We shall concentrate on the case of packets of fixed rate $R$ to simplify the subsequent discussion.
    This way, the only randomness in rate comes from the packet-erasure effect.
    Nevertheless, all the results that follow can be easily extended to random/varying rate allocations to which the effect of packet erasures $\{b_t\}$ is added in the same manner as in \eqref{eq:packet_erasures}.
\end{remark}

\begin{corol}[Distortion--rate region]
\label{corol:avg:IF:RDR:Dt}
    The distortion--rate region of Gauss--Markov multi-track with i.i.d.~$\Ber(\beta)$ packet erasures and instantaneous ACKs 
    is given 
    as in \thmref{thm:IF:RDR:Dt} with 
    \begin{align}
    \label{eq:B}
    	B \triangleq \E{2^{-2 \rone}} = 1 - \beta \left( 1 - 2^{-2 R} \right) .
    \end{align}
\end{corol}

\begin{corol}[Steady state]
\label{corol:InstantFB:SS}
	The steady-state distortion is given as in \colref{corol:RandomRate:SS} with 
    $B$ as in \eqref{eq:B}.
\end{corol}

\begin{remark}
	In contrast to the scenario without packet erasures, the uniform rate allocation can be improved by allowing a \textit{dynamic} rate allocation that \textit{depends on the pattern of packet erasures} $b^{t-1}$. 
	This setup can be thought of as the source-coding dual of the 
fast fading channel coding problem where the fading coefficient is known at both the transmitter and the receiver prior to transmission, 
	and the transmitter optimizes the transmission rate via waterfilling across time~\cite[Ch.~5.4]{TseViswanathBook}.
\end{remark}

%--------------------------------------------------------------------------------------------------------

\subsection{Multiple Packets Per Frame}
\label{ss:InstantFB:ManyPackets}

In \secref{ss:InstantFB:1packet} we assumed that one packet ($\PACKET_t$) was sent per each frame ($\bs_t$).
Instead, one may choose to transmit multiple packets of lower rate per one frame. 
If we assume that each packet arrival is instantly acknowledged, 
then the resulting scenario falls again in the random-rate budget framework of \secref{s:RandomRate}.
Interestingly, it turns out that the optimal number of packets per frame depends on the rate's PDF, \ie, increasing the number of packets can either improve or deteriorate the performance.

Specifically, assume that the observer uses $\Npackets$ packets of equal rate $R / \Npackets$ (and hence a total rate of $R$)
to successively refine~\cite[Ch.~13.5]{ElGamalKimBook} a single state frame $\bs_t$. Then, the rate probability distribution amounts to 
\begin{align}
	\rt &= \frac{b_t}{\Npackets} R ,
\end{align}
with $b_t$ denoting the number of successful packet arrivals at time $t$, corresponding to state frame $\bs_t$. Assuming that the erasure events of all packets are i.i.d.\ with probability \mbox{$1-\beta$} implies that $\{b_t\}$ are i.i.d.\ according to a Binomial distribution $\Bin \left(\Npackets, \beta \right)$.

\begin{figure}[t]
\centering
    \includegraphics[width=.88\columnwidth]{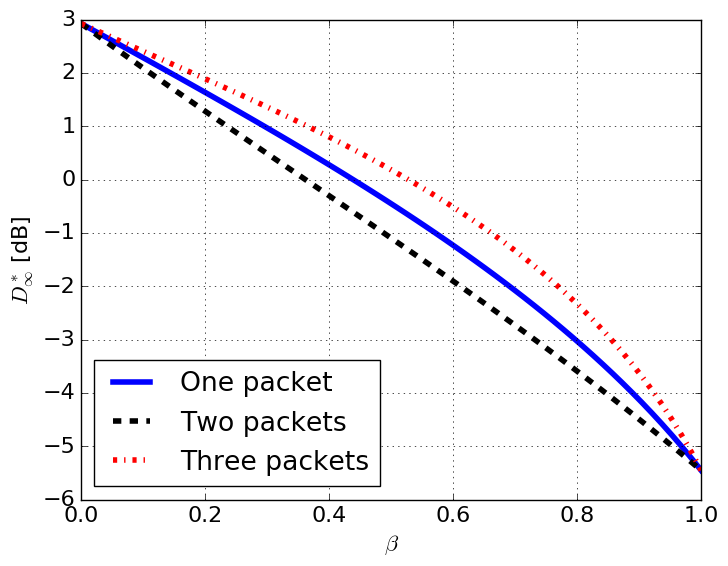}
    \caption{Evaluation of $D^*_\infty$ for $\Npackets = 1, 2$ and $3$ packets, all possible values of $\beta \in [0, 1]$, $R = 1$, $\alpha = 0.7$ and $W = 1$.}
\label{fig:multiple_packets}
\end{figure}

Interestingly, the optimal number of packets $\Npackets$ depends on the (total) rate $R$ and successful packet-arrival probability $\beta$, 
since by allocating more lower-rate packets, 
one trades a lower probability of receiving the maximal available rate at the state estimator with a higher probability of receiving intermediate rates. 
The optimal $\Npackets$ is determined by the number that minimizes $\E{2^{-\rt}}$, 
as is demonstrated in \figref{fig:multiple_packets}.

We note that in absence of ACKs of intermediate packets, the successive refinement encoding considered here cannot be used.
One could use repetition coding to trade multiplexing gain with diversity~\cite{TseViswanathBook} or multiple description coding~\cite{MD4Control:Jan:EURASIP}, when ACKs are sent only after all the intermediate packets are transmitted.
We do not discuss such extensions in this paper due to a lack of space.

\begin{remark}
	We only considered uniform rate allocations for all the packets. Clearly, one can generalize the same approach to non-uniform packet rates.
\end{remark}

\begin{remark}
	In practice one might expect longer packets to be prone to higher erasure probability. This can be taken into account when deciding on the $\Npackets$ that minimizes $\E{2^{-2 \rt}}$.
\end{remark}

%%%%%%%%%%%%%%%%%%%%%%%%%%%%%%%%%%%%%%%%%%%%%%%%%%%%%%%%%%%%%%%%%%%%%%%%%%%%%%%%%%%%%%%%%%%%%%%%%%%%%%%%%%

\section{Packet Erasures with Delayed ACKs}
\label{s:Kaspi}

% In this section 
We now tackle the case of i.i.d.\ packet erasures with ACKs that are delayed by one time unit, \ie, 
the case where at time $t$ the observer does not know whether the last packet arrived or not (\ie, it does not know $b_{t-1}$), but knows the erasure pattern of all preceding packets (knows $b^{t-2}$).
The observer~\eqref{eq:Encoder:Rule} and 
state estimator~\eqref{eq:hs_t:def} mappings can be written as [recall the definition of $g_t \triangleq b_t \PACKET_t$ in~\eqref{eq:received_packet}]:
\begin{align}
    \PACKET_\tind &= \mF_\tind \left( \bs^\tind, g^{\tind-2} \right) ,
 	\\* \hbs_\tind &= \mG_\tind \left( g^\tind \right) .
\end{align}

To construct a transmission scheme for this case, 
we recall the following result by Perron \etal~\cite[Th.~2]{PerronDiggaviTelatar:GaussianKaspi:ISIT2006}, 
which is a specialization to the jointly Gaussian case of the result by Kaspi~\cite[Th.~1]{Kaspi94}, who established the rate--distortion region of lossy compression with two-sided SI 
\begin{remark}
    Kaspi's result~\cite[Th.~1]{Kaspi94} can also be viewed as a special case of \cite{HeegardBerger85} with some adjustments; see~\cite{Causal-Noncausal_IEEEI2012}.
\end{remark}    

\begin{thm}[\!\!{\cite[Th.~2]{PerronDiggaviTelatar:GaussianKaspi:ISIT2006}}]
\label{thm:Kaspi}
    Let $\bs$ be an i.i.d.\ zero-mean Gaussian process of power $S$, 
    which is jointly Gaussian with SI $\by$, which is available at the observer 
    and satisfies $\bs = \by + \bz$, 
    where $\bz$ is an i.i.d.\ Gaussian noise of power~$Z$ that is independent of $\by$.
    Denote by $\hbs^+$ and $\hbs^-$ the reconstructions of $\bs$ with and without the SI $\by$, 
    and by $D^+$ and $D^-$~--- their mean squared error distortion requirements, respectively.
    Then, the smallest rate required to achieve these distortions is given by 
    \begin{align}
        &R^\mathrm{Kaspi}(S, Z, D^-, D^+)
    \\* 
        &\: = \begin{cases}
            0, & \IF D^- \geq S \AND D^+ \geq Z
	 \\ \RDF{S}{D^-}, & \IF D^- < S \AND D^+ \| S \geq D^- \| Z
	 \\ \RDF{Z}{D^+}, & \IF D^+ < Z \AND D^- \geq D^+ + S - Z
	 \\ \RDF{S}{D^- - \Delta^2}, & \IF \left\{\begin{array}{l} D^- < S \AND D^+ \| S < D^- \| Z \\ \AND D^- < D^+ + S - Z \end{array} \right.
        \end{cases}
  	\nonumber
  	\end{align}
    where 
    \mbox{$a \| b \triangleq \frac{a b}{a + b}$} denotes the harmonic mean of $a$ and $b$, 
    and 
    \begin{align}
     \Delta \triangleq& \frac{\sqrt{(S - Z) (S - D^-)} D^+ - \sqrt{(Z-D^+) ( D^- - D^+ )} S }{\sqrt{Z} \left( S - D^+ \right)} \,.
    \nonumber
    \end{align}
\end{thm}

\begin{remark}
\label{rem:SI@Tx_helps}
    Surprisingly, as observed by Perron \etal~\cite{PerronDiggaviTelatar:GaussianKaspi:ISIT2006}, 
    if the SI signal $\by$ is not available at the observer~--- 
    corresponding to the case considered in \cite[Th.~2]{Kaspi94}, \cite{HeegardBerger85}~---
    the required rate can be strictly higher than that in \thmref{thm:Kaspi}.
    This is in stark contrast to the case where the SI is not available at the observer, and the case where the SI is always available at the state estimator studied by Wyner and Ziv~\cite{WynerZiv76,Wyner78}. Knowing the SI at the observer allows to (anti-)correlate the noise $\bz$ with the quantization error~--- an operation that is not possible when the SI is not available at the observer, as the two noises must be independent in that case. This leads to some improvement, though a modest one, as implied by the dual channel-coding results~\cite[Prop.~1]{GaussianInput_ZamirErez03}, \cite{PhilosofZamir07}.
\end{remark}

In our case, at time $\tind$, the previous packet $\PACKET_{\tind-1}$ serves as the SI. Note that this SI is always available to the observer; the state estimator may or may not have access to it, depending whether the previous packet arrived or not. Since the ACK is delayed, during the transmission of the current packet $\PACKET_\tind$ the observer does not know whether the previous packet was lost.

The tradeoff between $D^+$ and $D^-$ for a given rate $R$ will be determined by the probability of a successful packet arrival~$\beta$.

\begin{scheme}[Kaspi-based]
\ 

    \emph{Observer.} At time $\tind$:
    \begin{itemize}
    \item
        Generates the prediction error 
        \begin{align}
            \label{eq:Kaspi:source2BeQuantized}
            \tbs_\tind \triangleq \bs_\tind - \alphat \alpha_{\tind-1} \hbs_{\tind-2} \,.
        \end{align}
    \item
        Generates $\PACKET_\tind$ by quantizing the prediction error $\tbs_\tind$
        as in \thmref{thm:Kaspi}, where $f_{\tind-1}$ is available as SI 
        at the observer and possibly at the state estimator (depending on $\betaut_{\tind-1}$)
        using the optimal quantizer of rate $R$ and frame length~$N$ that minimizes 
        the distortion
        averaged over $\betaut_{\tind-1}$:
        \begin{align}
        \label{eq:Kaspi:scheme:weightedD}
            \Dweight_\tind = \beta D^+_\tind + (1 - \beta) D^-_\tind \,;
        \end{align}
        more precisely, since the observer does not know $(\betaut_{\tind-1}, \betaut_\tind)$ at time $\tind$:
        \begin{itemize}
        \item
            Denote the reconstruction from $\PACKET_\tind$ and $g^{\tind-2}$~---
            namely given that $\betaut_\tind = 1$ and $\betaut_{\tind-1} = 0$~--- by  $Q^-_\tind(\tbs_\tind)$, 
            and the corresponding distortion by $D^-_\tind$.
        \item 
            Denote the reconstruction from $(\PACKET_{\tind-1}, \PACKET_\tind)$ and $g^{\tind-2}$~--- namely given that $\betaut_\tind = 1$ and $\betaut_{\tind-1} = 1$~--- by  $Q^+_\tind(\tbs_\tind)$, and the corresponding distortion by $D^+_\tind$.
        \item
            Denote the reconstruction of $\tbs_\tind$ at the state estimator from $\PACKET_\tind$ and $g^{\tind-1}$~--- namely given that \mbox{$\betaut_\tind=1$}~--- by $Q_\tind(\tbs_\tind)$, 
            and the corresponding distortion, averaged over $\betaut_{\tind-1}$, by $\Dweight_\tind$.
        \end{itemize}
        Then, the observer 
        sees $\alphat Q_{\tind-1}(\tbs_{\tind-1})$ as possible SI available at the state estimator to minimize $\Dweight_\tind$ as in \eqref{eq:Kaspi:scheme:weightedD}.
    \item
        Sends $\PACKET_\tind$ over the channel.
    \end{itemize}

\vspace{.1\baselineskip}
\emph{State estimator.} At time $\tind$:
    \begin{itemize}
    \item
	Receives $g_\tind$.
    \item
	Generates a reconstruction $\htbs_\tind$ of the prediction error $\tbs_\tind$:
    
    \begin{align}
	\label{eq:Kaspi:source_reconstruct}
        \htbs_t = 
        \begin{cases}
            Q^+_\tind(\tbs_\tind), & \betaut_\tind = 1, \betaut_{\tind-1} = 1
         \\ Q^-_\tind(\tbs_\tind), & \betaut_\tind = 1, \betaut_{\tind-1} = 0
         \\ 0,	         & \betaut_\tind = 0
        \end{cases}
    \end{align}
    \item
	Generates an estimate $\hbs_\tind$ of $\bs_\tind$:
	\begin{align}
	\label{eq:Kaspi:source_reconstruct:final}
	    \hbs_\tind = \alphat \hbs_{\tind-1} + \htbs_\tind \,.
	\end{align}
    \end{itemize}
\end{scheme}

This scheme is the optimal greedy scheme whose performance is stated next, in the limit of large $N$.
\begin{thm}
\label{thm:DelayedFB:UB}
% %    Let $\eps > 0$, however small. 
%     Then, for a large enough $N$, 
%     The expected distortion of the scheme at time \mbox{$\tind \in [2, T]$} given $(\betaut_1, \ldots, \betaut_\tind)$
%     satisfies the recursion
	The following distortions $D^T$ can be approached arbitrarily closely in the limit $N \to \infty$ for $\tind \in [2, T]$:
    \begin{subequations}    
    \label{eq:Kaspi:UB:Dt}
    \begin{align}
    	D_\tind &=
    	\begin{cases}
    	    D^+_\tind ,                  & \betaut_\tind = 1, \betaut_{\tind-1} = 1
	     \\ D^-_\tind ,                  & \betaut_\tind = 1, \betaut_{\tind-1} = 0
	     \\ \alphat^2 D_{\tind-1} + W , & \betaut_\tind = 0
    	\end{cases}
	\label{eq:Kaspi:UB:Dt:recursion}
    \\* D_1 &= D^+_1 = D^-_1 = \Wt 2^{-\betaut_1 2 R} + \eps ,
	\label{eq:Kaspi:UB:Dt:D0}
    \end{align}
    \end{subequations}
    where $D^+_\tind$ and $D^-_\tind$ are the distortions that minimize
    \begin{align}
    \label{eq:Kaspi:thm:weightedD}
        \Dweight_\tind = \beta D^+_\tind + (1 - \beta) D^-_\tind \,,
    \end{align}
    such that the rate of~\thmref{thm:Kaspi} satisfies 
    \begin{align}
        R^\mathrm{Kaspi}(\alphat D^-_{\tind-1} + W, \alphat D^+_{\tind-1} + W, D^-_\tind, D^+_\tind) = R .
    \end{align}
\end{thm}

\begin{IEEEproof}
	The proof is again the same as that of Ths.~\ref{thm:no-drop:RDR} and \ref{thm:IF:RDR:Dt}, with $\htbs_\tind$ generated as in \eqref{eq:Kaspi:source_reconstruct}. 
\end{IEEEproof}

\begin{remark}
    Here, in contrast to the case of instantaneous ACKs, evaluating the average distortions $\{D_t\}$ in explicit form (recall \corolref{corol:avg:IF:RDR:Dt}) is much more challenging. We do it numerically, instead.
\end{remark}

\begin{figure}[t]
    \includegraphics[width=\columnwidth]{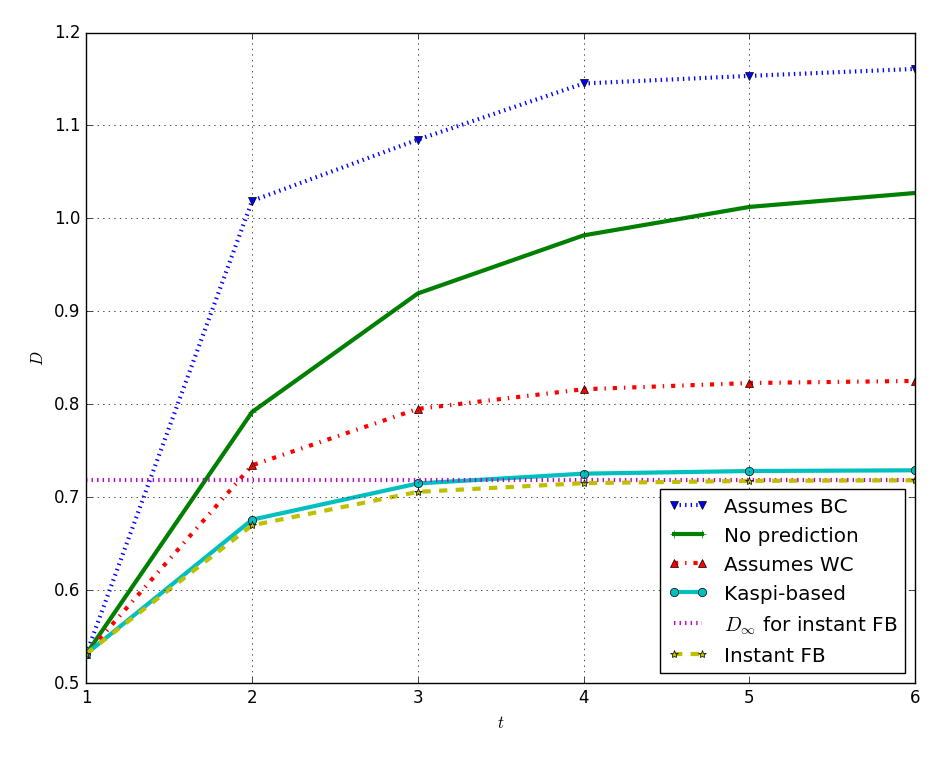}
\centering
\vspace{-2\baselineskip}
    \caption{Distortions $D_\tind$ as a function of the time $\tind$ 
    of the various schemes presented in this section, along with that of the instantaneous-ACK scheme of \secref{s:InstantFB}, 
    for $\alpha = 0.7$, $W = 1$, $\beta = 0.5$ and $R = 2$.}
\label{fig:performance}
\end{figure}

Somewhat surprisingly, the loss in performance of the Kaspi-based scheme due to the ACK delay is rather small compared to the scenario in \secref{s:InstantFB} where the ACKs are available instantaneously, for all values of $\beta$.\footnote{For $\beta$ values close to 0 or 1, the loss becomes even smaller as in these cases using the scheme of \secref{s:InstantFB} 
that assumes that the previous packet arrived or was erased, respectively, becomes optimal.} 
This is demonstrated in \figref{fig:performance}, where the perfomances of these schemes are compared along with the performances of
דthe following three simple schemes for $\alpha_t \equiv 0.7, W \equiv 1, \beta = 0.5, R = 2$ (we derive their performance for the special case of fixed parameters):

\begin{itemize}
\item \textbf{No prediction:}
    \begin{align}
        D_\tind &= \beta S_\tind 2^{-2R} + (1 - \beta) S_\tind , & \tind \in [T] \,,
    \end{align}
    where $S_\tind$ is the power of the entries of $\bs_\tind$ as given in~\eqref{eq:P}.
\item \textbf{Assumes worst case (WC):} 
    Since at time $\tind$ the observer does not know $b_{\tind-1}$, 
    a ``safe'' way would be to work as if $b_{\tind-1} = 0$. This achieves a distortion of 
    
    \begin{align}
    \\[-2.1\baselineskip]
        D_\tind &= \left[ \alpha^4 D_{\tind-2} + (1 + \alpha^2) W \right] \left[ \beta 2^{-2R} + (1-\beta)^2 \right]
        \\ &\qquad + \beta (1 - \beta) (\alpha^2 D_{\tind-1} + W), \qquad \tind = 2, \ldots, T \,,
    \nonumber
     \\ D_0 &= 0, \qquad\qquad D_1 = W 2^{-2R} .
     \nonumber
    \end{align}
\item \textbf{Assumes best case (BC):} 
    The optimistic counterpart of the previous scheme is that which always works as if $b_{\tind-1} = 1$. This scheme achieves a distortion of 
    \begin{align}
        &D_\tind = \beta \left[ \alpha^2 D_{\tind-1|\tind-2} 2^{-2R} + W \right) \left[ \beta 2^{-2R} + (1 - \beta) \right]
    \nonumber
    \\* &\qquad\quad + (1 - \beta) \left[ \alpha^2 D_{\tind-1|\tind-2} + W \right], \quad \tind = 2, \ldots, T \,,
    \nonumber
     \\* &D_{\tind-1|\tind-2} \triangleq \alpha^2 D_{\tind-2} + W, \hspace{2.15cm} \tind = 2, \ldots, T \,,
    \nonumber
     \\ &D_0 = 0, \quad D_1 = W 2^{-2R} .
    \end{align}
\end{itemize}

%%%%%%%%%%%%%%%%%%%%%%%%%%%%%%%%%%%%%%%%%%%%%%%%%%%%%%%%%%%%%%%%%%%%%%%%%%%%%%%%%%%%%%%%%%%

\section{Variable-Length Coding}
\label{s:variable_rate}

In contrast to previous sections where at time instant $t$ \emph{exactly} $N \Rt$ bits were available for the compression of the $N$-length vector $\bs_t$, 
in this section, we consider the less restrictive case, commonly referred to as VLC, where the (transmit) rate is constrained to $R$ only \emph{on average} across time~\cite{Huffman52}, \cite[Ch.~5]{CoverBook2Edition}. 
We assume again a packet-erasure case, where, 
as in \secref{ss:InstantFB:1packet}, 
the packet at time $t$ is erased with probability $1-\beta$, and successfully arrives with probability $\beta$. 
The packet-erasure events $\{b_t\}$ take values in $\{0,1\}$ where 0 corresponds to an erasure and 1~--- to a successful arrival; we assume that these events are i.i.d. 
We further concentrate on the scalar case, $N = 1$.
The rate constraint can be therefore written as:
\begin{align}
\label{eq:variableRate:constraint}
	\begin{aligned}
% 	\frac{1}{T} \sum_{t = 1}^T 
        \CE{\rt}{b_\tind = 1} &\leq R ,
     \\ \CE{\rt}{b_\tind = 0} &= 0 ,
	\end{aligned}
     && \tind \in [T] ,
\end{align}
where, in contrast to previos sections, in this section, \emph{$\rt$ can depend on the exact value of $\bs^t$}. 

\begin{remark}
	Similarly to the treatment in \secref{ss:InstantFB:ManyPackets}, 
	the treatment in this section can be extended to the case of multiple packets per state frame.
\end{remark}

% To make the results more meaningful, we shall limit ourselves to the fixed-budget (yet variable-rate) case case: $R_t \equiv R$.

We first note that the lower bound of \thmref{thm:IF:RDR:Dt} remains valid for the VLC case, 
since Shannon's classical rate--distortion theorem \cite{Shannon48,Shannon59,BergerBook} 
extends to the case of VLC (see, \eg, \cite{VariableLengthRDF:Hashimoto}).
We next prove that this lower bound can be closely met by incorporating 
ECDQ \cite{Ziv85,ZamirFeder:ECDQ,ZamirFeder:PrePostECDQ}, \cite[Ch.~5]{ZamirBook}, 
which is described as follows.

\begin{scheme}[ECDQ]
\label{scheme:ECDQ}
\

	\emph{Offline.}
%     \begin{itemize}
%     \item
    	The observer and the state estimator generate a common random dither $z$ that is uniformly distributed over $[-\Delta/2, \Delta/2)$.
%     \end{itemize}

	\emph{Observer.}
    \begin{itemize}
    \item
    	Uses a uniform-grid (one-dimensional lattice) quantizer with quantization step $\Delta$ to quantize $\gamma s + z$: \mbox{$Q_\Delta(\gamma s + z)$}, where $\gamma$ is a pre-determined scalar.
	\item
    	Applies entropy coding to the output of the quantizer.
	\item
    	Sends the output of the entropy coder.
    \end{itemize}

	\emph{State estimator.}
    \begin{itemize}
    \item
    	Receives the coded bits.
    \item
    	Reconstructs the output of the quantizer: $Q_\Delta(\gamma s + z)$.
	\item
    	Generates the state estimate by subtracting $z$ from the quantizer's output and multiplies the result by $\gamma$:
        \begin{align}
        	\hs = \gamma \left[ Q_\Delta(\gamma s + z) - z \right] .
        \end{align}
    \end{itemize}
\end{scheme}

\begin{thm}[ECDQ performance~{\cite{ZamirFeder:PrePostECDQ}, \cite[Ch.~5]{ZamirBook}}]
\label{thm:ECDQ}
	The average rate $\Recdq$ needed by the ECDQ scheme (for $N=1$) to achieve a distortion $\Decdq$ for a state $s$ with variance $S$ and $\gamma$ set to $\gamma = \sqrt{ 1 - D/S }$ 
    is bounded from above by 
	\begin{align}
    \label{eq:ECDQ:R:1D}
    	\Recdq \leq \half \log \frac{S}{\Decdq} + \half \log \frac{\twopi \e}{12} , 
    \end{align}
    where the first element in \eqref{eq:ECDQ:R:1D} is the Gaussian rate--distortion function 
    and the second element is the ``shaping loss''.
%     and the last element is the maximal rate-redundancy of a prefix free (entropy) code above the source entropy.
    
    Equivalently, the average distortion $\Decdq$ of the ECDQ scheme under an average rate constraint $\Recdq$~\eqref{eq:variableRate:constraint} is bounded from above by 
    \begin{align}
    \label{eq:ECDQ:D}
    	\Decdq \leq \frac{\twopi \e}{12} S 2^{-2 \Recdq} .
    \end{align}
\end{thm}

\begin{remark}[One-to-one source coding]
	The entropy coding employed here is assumed to be one-to-one, 
    that is, we do not require the resulting code to be prefix free.
    For a more thorough discussion of one-to-one versus prefix-free coding and 
    the rationale behind using each, see \secref{ss:discussion:1:1}.
\end{remark}

\begin{remark}[ECDQ for $N > 1$]
\label{rem:ECDQ:N>1}
	For $N > 1$, one may replace the uniform scalar quantizer with a lattice-based one;
    the resulting distortion in this case is upper bounded by 
	\begin{align}
    	\Recdq \leq \half \log \frac{S}{\Decdq} + \half \log \left( \twopi \e G_N \right) ,
    \end{align}
    where $G_N$ is the normalized second moment of the lattice. For the special case of a scalar lattice, 
    $G_1 = 1 / 12$. It is known, by the isoperimetric inequality \cite[Ch.~7]{ZamirBook}, that $G_N > \frac{1}{\twopi \e}$ for any lattice of any dimensions $N$. Moreover, it is known that a sequence of lattices of growing dimensions $N$ can be devised that attains this isoperimetric lower-bound in the limit of $N \to \infty$.
    See \cite{ZamirBook} for a thorough account of lattices and their application to ECDQ.
\end{remark}

We next incorporate ECDQ in the DPCM scheme of \secref{s:no-drops:IB}: we apply ECDQ (with i.i.d.\ dither $z_t$ across time) to $\ts_t$ to generate $\hts_t$ at the observer and recover it at the state estimator; 
the rest of the scheme remains exactly the same. 
We note that a similar scheme in the context of networked control (albeit without packet erasures) was previously proposed and analyzed in \cite{DerpichOstergaard:ECDQ4CausalRDF}.
The performance of \schemeref{scheme:ECDQ} is stated in the following theorem.

\begin{thm}[ECDQ-based DPCM scheme performance]
\label{thm:DPCM-ECDQ}
	The ECDQ-based DPCM scheme (for $N=1$) under an average rate constraint $\Recdq$~\eqref{eq:variableRate:constraint} achieves a distortion $\Decdq_t$ at time $t$ that satisfies the recursion:
    \begin{subequations}
    \label{eq:DPCM-ECDQ}
    \noeqref{eq:DPCM-ECDQ:recur,eq:DPCM-ECDQ:D0}
    \begin{align}
    	\Decdq_t &\leq \frac{\twopi \e}{12} B \left( \alphat^2 D_{t-1} + \Wt \right) ,
    \label{eq:DPCM-ECDQ:recur}
	\\* \Decdq_0 &= 0 ,
    \label{eq:DPCM-ECDQ:D0}
    \end{align}
    \end{subequations}
    with $\Decdq_0 = 0$ and
    $B$ as in~\eqref{eq:B}.
\end{thm}

This theorem suggests that the gap in performance of scalar systems compared to their $N$-dimensional counterparts 
is bounded by a multiplicative factor of $\twopi \e / 12$ in each recursive step~\eqref{eq:DPCM-ECDQ:recur}.

\begin{IEEEproof}
	The proof is identical to that in \secref{s:no-drops:IB}
    and of \thmref{thm:IF:RDR:Dt}, with 
    $D_t \leq (\alpha_t^2 D_{t-1} + W) B$ replaced with 
    $D_t \leq \frac{\twopi\e}{12} (\alpha_t^2 D_{t-1} + W) B$, 
    due to the shaping loss of ECDQ (recall \thmref{thm:ECDQ}).
\end{IEEEproof}

\begin{remark}[ECDQ-based DPCM scheme for $N > 1$]
	Following \remref{rem:ECDQ:N>1}, for the case of $N > 1$ the resulting distortion when applying ECDQ for $N > 1$ with an $N$-dimensional lattice is bounded from above by 
    \begin{align}
%     	\Decdq_t &\leq \twopi \e G_N \left( \alphat^2 D_{t-1} + \Wt \right) 2^{-2 \left( \Recdq - \frac{1}{N} \right)} ,
    	\Decdq_t &\leq \twopi \e G_N B \left( \alphat^2 D_{t-1} + \Wt \right) ,
	\\* \Decdq_0 &= 0 ,
    \end{align}
    where again $\Decdq_0 = 0$, $G_N$ is the normalized second moment of the lattice and $B$ is given in~\eqref{eq:B}.
\end{remark}

In the limit of large $T$, we attain the following steady-state distortion.

\begin{corol}[ECDQ-based DPCM scheme in steady-state]
\label{corol:DPCM-ECDQ:SS}
	If $\frac{\twopi \e}{12} \alpha^2 B < 1$, 
	then the steady-state distortion of the ECDQ-based DPCM scheme (for $N=1$) under an average rate constraint $\Recdq$~\eqref{eq:variableRate:constraint} is bounded from above by
    \begin{align}
    \label{eq:DPCM-ECDQ:D:SS}
    	\Decdq_\infty &\leq \frac{\frac{\twopi \e}{12} W B}{1 - \frac{\twopi \e}{12}\alpha^2 B} \,
    \end{align}
    where $B$ is given in~\eqref{eq:B}.
\end{corol}

\begin{remark}[Stabilizability]
	The stabilizability condition $\frac{\twopi \e}{12} \alpha^2 B < 1$ is distant from that of the case of large frames by the shaping loss $\frac{\twopi \e}{12}$. 
    This can be alliviated by applying downsampling, \ie, 
    sending $\kappa R$ bits (on average) every $\kappa \in \nats$ samples and remaining silent during the rest; 
    the resulting stabilizability condtion in this case becomes 
    $\sqrt[k]{\frac{\twopi \e}{12}} \alpha^2 B < 1$.
\end{remark}

%%%%%%%%%%%%%%%%%%%%%%%%%%%%%%%%%%%%%%%%%%%%%%%%%%%%%%%%%%%%%%%%%%%%%%%%%%%%%%%%%%%%%%%%%%%

\section{Application to Networked Control}
\label{s:control}

An important application of state tracking is to networked control, namely, 
to the scenario where, in contrast to traditional control, 
the observer is not co-located with the controller, 
and communicates with it instead via a noiseless (``packeted'') channel. 
Hence, the controller assumes the additional role of the state estimator.

We concentrate on the following simple setting, also depicted in \figref{fig:control}.
The channel is the noiseless random-rate budget channel of \secref{s:RandomRate}.

We consider a stochastic system with discrete-time linear scalar plant evolution which is the same as in~\eqref{eq:s_t:recursion}:
\begin{subequations}
\label{eq:plant}
\noeqref{eq:plant:recur,eq:plant:X0}
\begin{align}
	s_t &= \alpha s_{t-1} + w_t + u_{t-1} ,
\label{eq:plant:recur}
 	& t \in [T]
\\* s_0 &= 0 ,
\label{eq:plant:X0}
\end{align}
\end{subequations}
where the coefficient $\alpha$ (which is usually assumed to be fixed across time in control applications) can be greater than 1 in its absolute value, corresponding to an unstable open-loop process, 
with the additional term $u_{t-1}$ serving as the control action that is generated by the controller from all past packets $\PACKET^{t-1}$, 
and is used to stabilize the system. 

We consider the random-rate budget scenario of~\secref{s:RandomRate}.
The goal of the system is to minimize the average-stage LQG cost upon reaching the horizon~$T$:
\begin{align}
\label{eq:cost}
	\oLQGcost_T &\triangleq
    \frac{1}{T}\E{ \sum_{t = 1}^{T-1} \left( \CostXs_t s_t^2  + \CostUs_t u_t^2 \right) + \CostLastX s_T^2 } ,
\end{align}
where $\{\CostXs_t\}$ and $\{\CostUs_t\}$ are known non-negative scalars, respectively, that penalize the cost for state deviations and control actuations, respectively. 

In order to derive bounds on the LQG cost for this setting, 
we use a result by Fischer~\cite{QuantizedLQG:Fischer82} and by~Tatikonda \etal~\cite{TatikondaSahaiMitter},
that extends the celebrated control-theoretic separation principle 
to networked control systems. 

\usetikzlibrary{shapes,arrows,decorations.markings}

\tikzstyle{plant} = [draw, fill=red!5, rectangle, 
    minimum height=3em, minimum width=6em]
\tikzstyle{block} = [draw, fill=blue!5, rectangle, 
    minimum height=3em, minimum width=6em]
\tikzstyle{sum} = [draw, fill=yellow!10, circle, node distance=1cm]
\tikzstyle{coord} = [coordinate]
\tikzstyle{gain} = [draw, fill=red!5, regular polygon, regular polygon sides=3, shape border rotate=-90]
\tikzstyle{pinstyle} = [pin edge={to-,thick,black}]

\tikzstyle{BitPipe} = [thick, decoration={markings,mark=at position
   1 with {\arrow[semithick]{open triangle 60}}},
   double distance=1.4pt, shorten >= 5.5pt,
   preaction = {decorate},
   postaction = {draw,line width=1.4pt, white,shorten >= 4.5pt}]

\begin{figure}
\centering
	\scalebox{.85}{
    \begin{tikzpicture}[auto, arrow/.style={very thick, ->, >=stealth'},node distance=.2\columnwidth,>=latex']
        \node [coord] (input) {};
        \node [plant, right of = input, node distance=.36\columnwidth] (plant) {$s_t = \alpha s_{t-1} + w_t + u_{t-1}$};
        \node [coord, right = of plant] (midoutput) {};
        \node [coord, right = of midoutput] (output) {};

        \node [block, below of=midoutput, node distance = .2\columnwidth] (enc) {Observer};
        \node [block, below of=plant] (dec) {Controller};

        \draw[arrow] (input) -- node {$w_t$} (plant);
        \draw[arrow] (plant) -- node (yArrow) {$s_t$} (output);
        \draw[arrow] (yArrow) -- node[above] {} (enc);
        \draw[BitPipe] (enc) -- node[align=center] {$R_t$ \\ bits} (dec);
        \draw[arrow] (dec) -- node {$u_t$} (plant);
    \end{tikzpicture}}
    \caption{Linear control system with a finite-rate feedback.}
    \label{fig:control}
\end{figure}
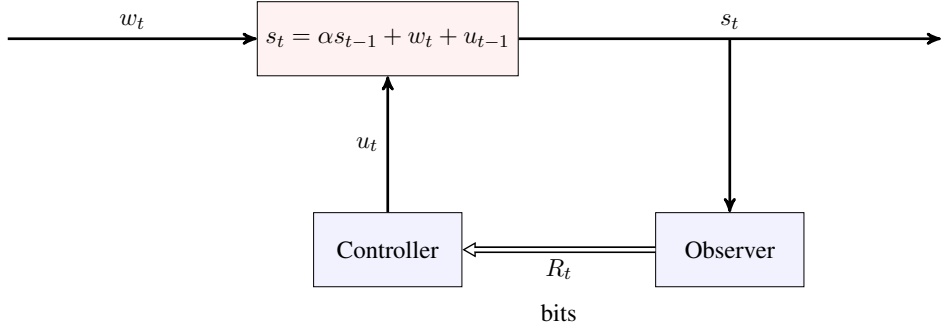

\begin{lemma}[\!\!{\cite{QuantizedLQG:Fischer82,TatikondaSahaiMitter}}]
\label{lem:scalar:separation}
    The optimal controller
	is given by 
    \begin{align}
    \label{eq:lem:scalar:separation:controller}
    	u_t &= -K_t  \hs_t , 
    \end{align}
    where $\hs_t \triangleq \CE{s_t}{\PACKET^t}$, $K_t$ is the optimal linear quadratic regulator (LQR) control gain
    \begin{align}
    \label{eq:LQG:scalar:Kt}
		K_t &= \frac{L_{t+1}}{\CostUs_t + L_{t+1}} \alpha ,
    \end{align}
    and $L_t$ satisfies the dynamic backward Riccati recursion~\cite{BertsekasControlVol1}:
    \begin{align}
    \label{eq:lem:LQG:Lt}
    	L_t &= \CostXs_t + \alpha \CostUs_t K_t  ,
    \end{align}
    with $L_{T+1} = 0$.\footnote{In case $\CostUs_T = 0$, define $K_T = 0$.}
	Moreover, this controller achieves a cost of 
    \begin{align}
    \label{eq:LQG:cost}
    	\oLQGcost_T = \frac{1}{T} \sum_{t = 1}^T \left\{ W L_t + \alpha K_t L_{t+1} \E{\left(s_t - \hs_t \right)^2} \right\} ,
    \end{align}
    where we use the convention $R_T = 0$ and $\PACKET_T = 0$ for the definition of $\hs_T$, as no transmission or control action are performed at time $T$.
\end{lemma}

%----------------------------------------------------------------------------

\subsection{Lower Bound}

By substituting the result of \thmref{thm:IF:RDR:Dt} into \lemref{lem:scalar:separation}, we 
attain the following lower bound for the achievable LQG cost,
which extends the result of~\cite{KostinaHassibi:RDF4Control:AC} to the case of random-rate budgets.

\begin{thm}[LQG cost lower bound]
\label{col:cost:LB:scalar}
	The optimal LQG cost~\eqref{eq:cost} with rate tuple $R^T$ 
    is bounded from below by 
    \begin{align}
    \label{eq:LQG:cost:LB}
    	\oLQGcost_T \geq \frac{1}{T} \sum_{t = 1}^T \left\{ W L_t + \alpha K_t L_{t+1} D^*_t \right\} ,
    \end{align}
    where $K_t$ and $L_t$ are defined as in \lemref{lem:scalar:separation}, 
    and $D^*_t$~--- in~\eqref{eq:RDR:IF:Dt}. 
\end{thm}

\begin{IEEEproof}
	The proof is immediate by noting that, similar to \eqref{eq:general:UB:errors_relation}, 
    at time $t$, given $\PACKET^t$, all the past control actions $u^{t-1}$~--- being a deterministic function of $\PACKET^{t-1}$~--- can be absorbed into~$\hs_t$.
\end{IEEEproof}

%----------------------------------------------------------------------------

\subsection{Variable-Length Coding}

Similarly to the proof of \thmref{col:cost:LB:scalar},by combining the results of \thmref{thm:DPCM-ECDQ} and \lemref{lem:scalar:separation} we attain the following upper bound for the achievable LQG cost, in the VLC scenario; following the exposition in \secref{s:variable_rate}, we concentrate here on the packet-erasure channel.
\begin{thm}[VLC LQG cost upper bound]
\label{col:cost:UB:scalar}
	The LQG cost~\eqref{eq:cost} for the VLC scenarios under an average-rate constraint 
    $R$ \eqref{eq:variableRate:constraint}, is bounded from above by 
    \begin{align}
    \label{eq:LQG:cost:LB}
    	\oLQGcost_T \leq \frac{1}{T} \sum_{t = 1}^T \left\{ W L_t + \alpha K_t L_{t+1} \Decdq_t \right\} ,
    \end{align}
    where $K_t$ and $L_t$ are given in \lemref{lem:scalar:separation}, 
    and $\Decdq_t$ is bounded from above as in~\eqref{eq:DPCM-ECDQ:D:SS}.
\end{thm}

\begin{IEEEproof}
	Again, the proof is immediate by noting that, similar to the impossibility proof of~\secref{s:no-drops:OB}, 
    at time $t$, given $\PACKET^t$, all the past control actions $u^{t-1}$~--- being a deterministic function of $\PACKET^{t-1}$~--- are fully determined.
\end{IEEEproof}

%----------------------------------------------------------------------------

\subsection{Steady State}

We consider here the fixed-parameter fixed-rate case:
\begin{subequations}
\label{eq:LQG:fixedParams}
\noeqref{eq:LQG:fixedParams:Q,eq:LQG:fixedParams:R,eq:LQG:fixedParams:rate}
\begin{align}
	\CostXs_t &\equiv \CostXs ,
\label{eq:LQG:fixedParams:Q}
 \\ \CostUs_t &\equiv \CostUs ,
\label{eq:LQG:fixedParams:R}
 \\ \Rt &\equiv R , 
\label{eq:LQG:fixedParams:rate}
\end{align}
\end{subequations}
and similarly to the steady-state distortion \eqref{eq:D:SS} and average-stage steady-state distortion~\eqref{eq:totalRateBudgetConstraint:SS}, 
we wish to determine the optimal steady-state average-stage cost
\begin{align}
\label{eq:cost:SS}
    \oLQGcost_\infty \triangleq \limsup_{T \to \infty} \oLQGcost_T \,.
\end{align}

\begin{corol}[LQG cost lower bound]
\label{col:cost:LB:scalar:SS}
	The steady-state LQG cost for the fixed-parameter fixed-rate case~\eqref{eq:LQG:fixedParams} is bounded from below by 
    \begin{align}
    \label{eq:LQG:cost:LB}
    	\oLQGcost_\infty \geq W L_\infty + \alpha K_\infty L_\infty D^*_\infty \,,
    \end{align}
    where $D^*_\infty$ is given in~\eqref{eq:RandomRate:SS:D}, 
    \begin{align}
    \label{eq:Kinf}
    	K_\infty = \frac{L_\infty}{\CostUs + L_\infty} \alpha ,
    \end{align}
    and $L_\infty$ is the positive solution of 
    \begin{align}
    \label{eq:Linf}
    	L_\infty^2 - \left[ \left( \alpha^2 - 1 \right) \CostUs + \CostXs \right] L_\infty - \CostXs \CostUs = 0.
    \end{align}
\end{corol}

\begin{remark}[Fixed- versus variable-length coding]
	As noted in \secref{s:variable_rate}, the result of \colref{col:cost:LB:scalar:SS} holds true for VLC and hence also for the more restrictive FLC.
\end{remark}

\begin{remark}[Comparison to separation-based bounds]
	In~\cite{KostinaHassibi:RDF4Control:AC}, 
    it is shown that the optimal steady-state LQG cost must satisfy~\eqref{eq:LQG:cost:LB}
    with the distortion $D^*_\infty$ dictated by the source--channel separation between
    the \textit{causal rate--distortion} $R_\text{C}(D_\infty)$~\cite{GorbunovPinsker:CausalRDF,TatikondaPhD}
    and the directed capacity (maximal directed information)~\cite{Massey:DirectedInfo}.
    Since in our case the directed capacity is upper bounded by the regular capacity of the channel,
    $C = \E{\rone}$, and the causal rate--distortion function (which is in itself a lower bound) is given by~\cite{GorbunovPinsker:CausalRDF:Gauss,TatikondaPhD} 
    \begin{align}
    	R_\text{C}(D^*_\infty) = \half \log \left( \alpha^2 + \frac{W}{D^*_\infty} \right) ,
    \end{align}
    the source--channel separation-based bound $R_\text{C}(D^*_\infty) \leq C$ 
    reduces to the expression in~\eqref{eq:RandomRate:SS:D}
    with $B \triangleq \E{2^{-2\rone}}$ replaced with $B_\text{Sep} \triangleq 2^{-2\E{\rone}}$.
    By applying Jensen's inequality we see that $B < B_\text{Sep}$ for any non-deterministic rate budget distirubtion. 
    Thus, the joint source and channel treatment offered in this work strengthens the separation-based adaptation of the results in~\cite{KostinaHassibi:RDF4Control:AC}.
    The difference becomes especially pronounced in the packet-erasure and instantaneous ACKs scenario of \secref{ss:InstantFB:1packet} with an infinite transmission rate $R$ [recall \eqref{eq:packet_erasures}]~--- a setting extensively studied in the past decade~\cite{Sinopoli:IntermittentObservations,GuptaDanaHespanhaMurrayHassibi_EstimationControl,SchenatoSinopoliFranceschettiPoolaSSS}.
    In this case, $B_\text{Sep}$, and consequently also the lower bound on $D^*_\infty$, reduces to the trivial zero bound, whereas $B = 1 - \beta > 0$ unless $\beta = 1$.
\end{remark}

\begin{corol}[VLC LQG cost upper bound]
\label{col:cost:UB:scalar:SS}
	The steady-state LQG cost for the packet-erasure fixed-parameter case under an average rate constraint $R$~\eqref{eq:variableRate:constraint} is bounded from above by    
    \begin{align}
    \label{eq:LQG:cost:LB}
    	\oLQGcost_\infty \leq W L_\infty + \alpha K_\infty L_\infty \Decdq_\infty \,,
    \end{align}
    where $\Decdq_\infty$, $K_\infty$, $L_\infty$ are given in~\eqref{eq:DPCM-ECDQ:D:SS}, \eqref{eq:Linf}, \eqref{eq:Kinf}, respectively.
\end{corol}
%%%%%%%%%%%%%%%%%%%%%%%%%%%%%%%%%%%%%%%%%%%%%%%%%%%%%%%%%%%%%%%%%%%%%%%%%%%%%%%%%%%%%%%%%%%

\section{Discussion}
\label{s:discussion}

%------------------------------------------------------------------------------------------
\subsection{ACKs with Larger Delays}

To extend the delayed ACK scheme of \secref{s:Kaspi} for the case of delayed ACKs by one time instant,
to larger delays, a generalization of \thmref{thm:Kaspi} is needed.
Unfortunately, the optimal rate--distortion region for more than two SI options (\eg, with or without~correlated SI $\by$) remains an open problem and is only known for the (``degraded'') case when 
the state and the possible SIs form a Markov chain.
Nonetheless, achievable regions for multiple SI options have been proposed in \cite{HeegardBerger85}, 
which can be used for the construction of schemes that accommodate larger delays.

%------------------------------------------------------------------------------------------
\subsection{Scalar Fixed-length Coding}

In this paper we derived lower bounds and proved that they are tight 
in the limit of large values of $N$.
In the case of scalar FLC quantization, 
both design and analysis of good schemes are more involved and remain beyond the scope of this paper. 
For a treatment
of the case of logarithmically concave noise distributions (Gaussian included), 
see~\cite{FixedRateLQG:CDC2017}.

%------------------------------------------------------------------------------------------
\subsection{Prefix-Free Versus One-Shot Lossless Compression}
\label{ss:discussion:1:1}

The VLC ECDQ-based schemes throughout this work employed one-to-one lossless coding.
This is a reasonable assumption since, 
in packeted communications, the descriptions of subsequent symbols may be assumed to be parsed by the underlying protocol, which allows, in turn, to part with the prefix-free constraint and attain better performance~\cite{AlonOrlitsky:1to1Compression}.
Specifically, 
the bit loss with respect to the entropy of the process of prefix-free coding is  circumvented by one-to-one coding~\cite{Wyner:1to1Compression}.
Nonetheless, the results of this paper can be easily adjusted to the prefix-free coding case 
by adding an extra bit on the right hand side of~\eqref{eq:ECDQ:R:1D}~--- the maximal loss of prefix-free entropy coding above the entropy, and replacing the factor $\twopi \e / 12$ in \eqref{eq:ECDQ:D}--\eqref{eq:DPCM-ECDQ:D:SS} by $\twopi \e / 3$.

%------------------------------------------------------------------------------------------
\subsection{Packet-Erasure Modeling}
\label{ss:discussion:packet-drop-model}

In this work, we modeled the packet erasures by an i.i.d.\ process. 
Nonetheless, the derived results can be extended far beyond this setting, as is evident from the proof of \thmref{thm:IF:RDR:Dt}.

In the VLC setting, the erasure probability is likely to be higher for longer packets, and calls for further investigation.

%------------------------------------------------------------------------------------------
\subsection{Non-Gaussian}
\label{ss:discussion:nonGauss}

Following \assertref{thm:LB:nonGaussian}, the lower bounds in this work can be extended to the 
case of a non-Gaussian driving process $\bw_t$, in a straightforward fashion.

%%%%%%%%%%%%%%%%%%%%%%%%%%%%%%%%%%%%%%%%%%%%%%%%%%%%%%%%%%%%%%%%%%%%%%%%%%%%%%%%%%%%%%%%%%%

\bibliographystyle{IEEEtran}

%%%%%%%%%%%%%%%%%%%%%%%%%%%%%%%%%%%%%%%%%%%%%%%%%%%%%%%%%%%%%%%%%%%%%%%%%%%%%%%%%%

%%%%%%%%%%%%%%%%%%%%%%%%%%%%%%%%%%%%%%%%%%%%%%%%%%%%%%%%%%
%%%%%%%%%%%%%%%%%%%%%%%%%%%%%%%%%%%%%%%%%%%%%%%%%%%%%%%%%%
%%%%%%%%%%%%%%%%%%%%%%%%%%%%%%%%%%%%%%%%%%%%%%%%%%%%%%%%%%
\end{document}